\documentclass{article}

 \usepackage[utf8]{inputenc}
 \usepackage[english]{babel}
 \usepackage{cite}
 \usepackage{amsfonts}
 \usepackage{graphicx, epstopdf}
 \usepackage[useregional]{datetime2}
 \usepackage{graphics}
 \usepackage{amsthm}
 \usepackage{amssymb}
 \usepackage{amsmath}
 \usepackage{caption}
 \usepackage{bbm}
 \usepackage[section]{placeins}
\usepackage{algorithm}
\usepackage{algpseudocode}
 \usepackage{subfig}
 \usepackage{xcolor}
 \usepackage{hyperref}
\usepackage{booktabs}
 \usepackage{array}
\usepackage{verbatim}
\usepackage{float}
\usepackage{placeins}
\usepackage{stackengine}
\usepackage{multirow}
\usepackage[tablename=TABLE]{caption}
\usepackage[figurename=FIG.]{caption}

\DeclareMathOperator{\diam}{diam}

\DeclareMathOperator{\infinitesmal}{d}

\def\d{\mathbbm{d}}

\newtheorem{theorem}{Theorem}[section]
\newtheorem{corollary}{Corollary}[section]
\newtheorem{lemma}{Lemma}[section]
\newtheorem{proposition}{Proposition}[section]

\theoremstyle{definition}
\newtheorem{definition}{Definition}[section]

\theoremstyle{definition}

\theoremstyle{remark}

\newcommand{\R}{\mathbb{R}} 
\newcommand{\N}{\mathbb{N}} 
\newcommand{\W}{\mathbb{W}} 
\newcommand{\D}{\mathcal{D}} 
\newcommand{\E}{\mathbb{E}} 
\newcommand{\Pro}{\mathbb{P}} 
\newcommand{\DxobsDy}{\mathcal{D}_{X_{O}}|D_{Y^{1:m}}} 
\newcommand{\1}{\mathbbm{1}}

\title{A Bayesian Framework for Persistent Homology }

\author{Vasileios Maroulas \qquad Farzana Nasrin \qquad Christopher Oballe \\ University of Tennessee, Knoxville }
\date{}
\begin{document}
\maketitle
\providecommand{\keywords}[1]{\textbf{\textit{Keywords}} #1}

\begin{abstract}
Persistence diagrams offer a way to summarize topological and geometric properties latent in datasets. While several methods have been developed that utilize persistence diagrams in statistical inference, a full Bayesian treatment remains absent. This paper, relying on the theory of point processes, presents a Bayesian framework for inference with persistence diagrams relying on a substitution likelihood argument. 
In essence, we model persistence diagrams as Poisson point processes with prior intensities and compute posterior intensities by adopting techniques from the theory of marked point processes. We then propose a family of conjugate prior intensities via Gaussian mixtures to obtain a closed form of the posterior intensity. Finally we demonstrate the utility of this Bayesian framework with a classification problem in materials science using Bayes factors.

\end{abstract}


\section{Introduction}
\begin{figure}[h!]
    \centering
     \subfloat[]{\includegraphics[width=2.2in,height=1.8in]{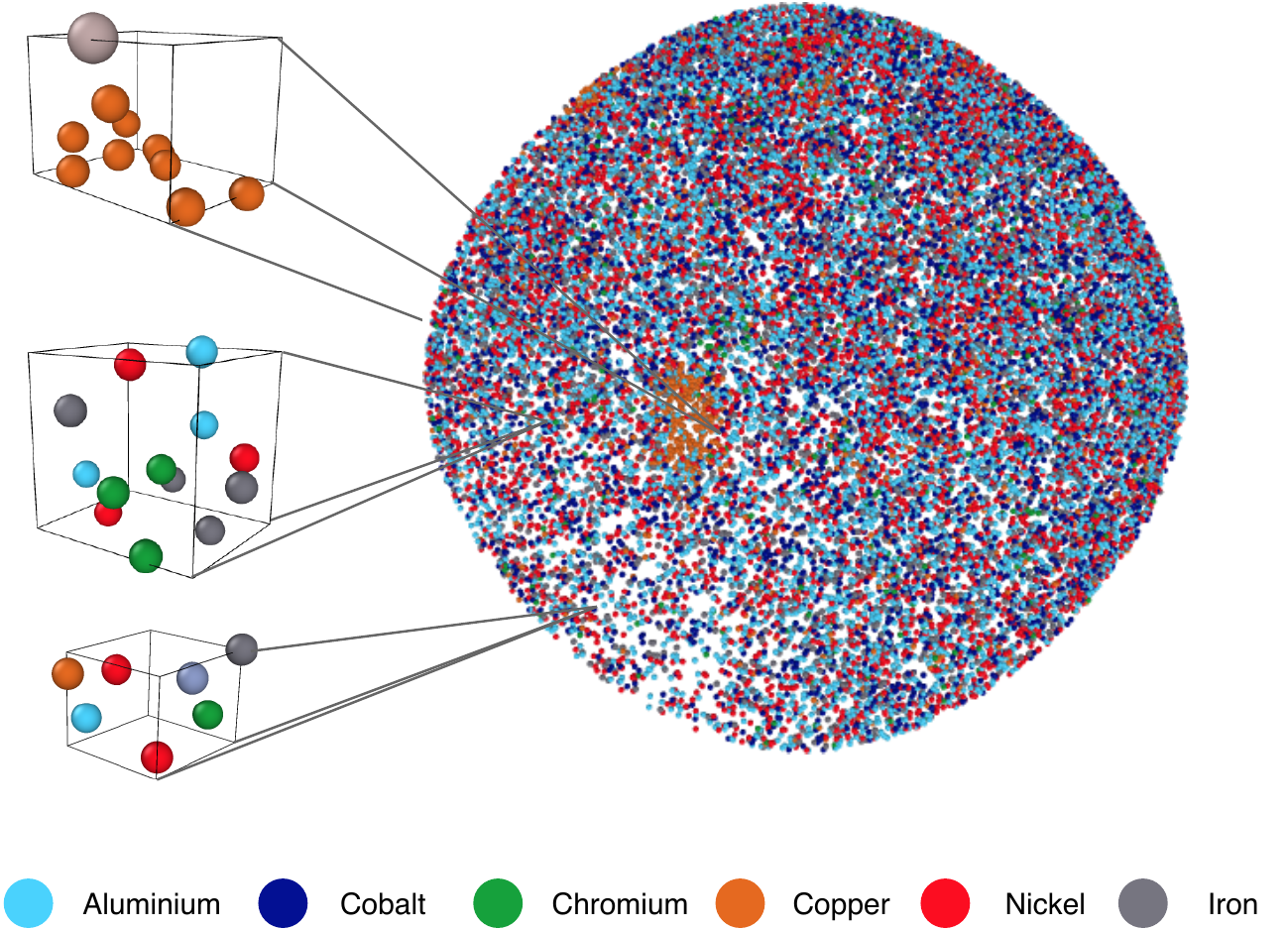}}\hfill
       \subfloat[]{\includegraphics[width=1.2in,height=1.4in]{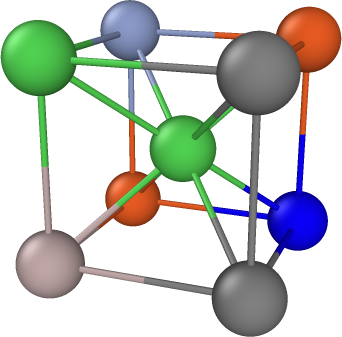}}\hfill
     \subfloat[]{\includegraphics[width=1.2in,height=1.4in]{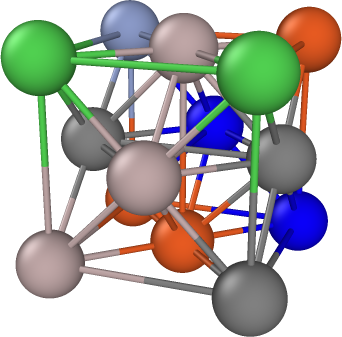}}\hfill
        \caption{\textit{(a) Image of APT data with atomic neighborhoods shown
            in detail on the left. Each
            pixel represents a different atom, the neighborhood of which is considered. Certain patterns with distinct crystal structures exist, e.g., the orange region
            is copper-rich (left), but overall no pattern is identified.
            Putting a single atomic cubic unit cell under a
            microscope, the true crystal structure of
            the material, which could be either body-centered cubic
              (BCC) (b) or  face-centered cubic (FCC) (c), is not revealed.
            This distinction is obscured due to further experimental noise. Notice there is 
        an essential topological difference between the two structures in (b) and (c):
        The BCC structure has one atom at its center, whereas the FCC
            is hollow in its center, but has one atom in the center of each of its faces.}}\label{fig:cells}
    \end{figure}

A crucial first step in understanding patterns and properties of a crystalline material is determining its
    crystal structure. For highly disordered metallic alloys, such as high entropy alloys (HEAs),
    atom probe tomography (APT) gives a snapshot of the local atomic environment; see Figure \ref{fig:cells}. However,
    APT has two main drawbacks: experimental noise and an abundance of missing data. Approximately 65\%
    of the atoms in a sample are not registered in a typical experiment \cite{santodonato2015deviation}, and the spatial coordinates of those identified atoms are corrupted by experimental noise \cite{miller2012future}. Understanding the atomic pattern within HEAs using an APT image requires observation of atomic cubic unit neighborhood cells under a microscope. This is problematic as APT may have a spatial 
    resolution approximately the length of the unit cell
    under consideration \cite{kelly2013atomic,miller2012future}. Hence, the process is unable to see the finer details of a material, rendering the
    determination of a lattice structure a challenging problem \cite{spannaus,mcnutt2017interfacial}. Existing algorithms for detecting the crystal 
    structure 
    \cite{chisholm2004new,hicks2018aflow,honeycutt1987molecular,larsen2016robust,moody2011lattice,togo2018spglib} are not able to
    establish the crystal lattice of an APT dataset,
    as they rely on symmetry arguments based on identifying repeating parts of molecules. Consequently, the field of atom probe crystallography, i.e., 
    determining the crystal
    structure from APT data, has emerged in recent years \cite{gault2012atom} and \cite{moody2011lattice}. Algorithms in this field rely
    on knowing the global lattice structure \emph{a priori} and aim to
    determine local small-scale structures within a larger sample. For some
    materials this information is readily known, while for others, such
    as HEAs, the global structure is unknown and must be inferred.
    
 A recent work \cite{ziletti2018insightful} proposes a machine-learning approach to classifying crystal structures
    of a noisy and sparse materials dataset without 
    knowing the global structure \emph{a priori}.
    The authors employ a convolutional
    neural network for classifying the crystal structure by examining
    a diffraction image, a computer-generated diffraction pattern.
    The authors suggest their method could be used to determine
    the crystal structure of APT data. However, the synthetic data
    considered in \cite{ziletti2018insightful} is not a realistic
    representation of experimental APT data, where about 65\% of the data is missing and furthermore corrupted by observational noise. Most importantly, their synthetic data is either sparse or noisy, not a combination of both. Herein, we consider a combination of noise and sparsity such as is the case in real APT data.
    
In this work, we specifically classify unit cells that are either body-centered cubic (BCC) 
    or face-centered cubic (FCC). These lattice structures are the 
    essential building blocks of HEAs \cite{zhang2014microstructures} and have
    fundamental differences that set them apart in the case of noise-free, complete materials data.
    The BCC structure has
    a single atom in the center of the cube, while the FCC has a void in its
    center but has atoms on the center of the cubes' faces, see Figure \ref{fig:cells} (b-c). These two crystal structures are distinct when viewed through the lens of topology. Differentiating between
    the empty space and connectedness of these two lattice structures allows us to create an
    accurate classification rule.
    This fundamental distinction between BCC and FCC point clouds
    is captured well by topological methods
    and explains the high degree of accuracy
    in the classification scheme presented herein. Indeed, we offer a Bayesian classification framework for persistence homology.

Overall, topological data analysis (TDA) encompasses a broad set of techniques that explore topological structure in datasets \cite{edelsbrunner2010computational, Ghrist07barcodes, tda_overview, Wasserman2018}. One of these techniques, persistent homology, associates shapes to data and summarizes salient features with persistence diagrams -- multisets of points that represent homological features along with their appearance and disappearance scales \cite{edelsbrunner2010computational}. Features of a persistence diagram that exhibit long persistence describe global topological properties in the underlying dataset, while those with shorter persistence encode information about local geometry and/or noise. Hence, persistence diagrams can be considered multiscale summaries of data's shape. 
 While there are several methods present in literature to compute persistence diagrams, we adopt geometric complexes that are typically used for applications of persistent homology to data analysis in various settings such as handwriting analysis \cite{Adcock2016}, studying of brain arteries \cite{bendich2016,Biscio2019}, image analysis \cite{Bonis2016,carriere2015,Carlsson2008}, neuroscience \cite{Chung2015,Sizemore2018,Babichev2017}, sensor network \cite{Dlotko2012,DeSilva2007}, protein structure \cite{Gameiro2015,Kusano2016}, biology \cite{Sgouralis2017,Mike2016,Nicolau2011}, dynamical system \cite{Khasawneh2016}, action recognition \cite{Venkataraman2016}, signal analysis \cite{Marchese2018,andy_dpc,Pereira2015,MMM}, chemistry \cite{Xia2015}, genetics \cite{Humphreys2019}, object data \cite{Patrangenaru2018}, etc.
 

 Researchers desire to utilize persistence diagrams for inference and classification problems. Several achieve this directly with persistence diagrams  \cite{Mike2018,Marchese2018,Bobrowski2017,pd_confidence_ints,Mileyko2011,Robinson2017,Bubenik2018}, while others elect to first map them into a Hilbert space \cite{pers_landscapes,multiscale_kernel,1507.06217,Turner2014,Fabio2015}.
The latter approach enables one to adopt traditional machine learning and statistical tools such as principal component analysis, random forests, support vector machines, and more general kernel-based learning schemes. Despite progress toward statistical inference, to the best of our knowledge, a full Bayesian treatment predicated upon creating posterior distributions of persistence diagrams is still absent in literature. The first Bayesian considerations in a TDA context take place in \cite{Mileyko2011} where the authors discuss a conditional probability setting on persistence diagrams where the likelihood for the observed point cloud has been substituted by the likelihood for its associated topological summary.

 The homological features in persistence diagrams have no intrinsic order implying they are random sets as opposed to random vectors. This viewpoint is embraced in \cite{Mike2018} to construct a kernel density estimator for persistence diagrams. This kernel density estimator gives a sensible way to obtain priors for distributions of persistence diagrams; however, 
 computing posteriors entirely through the random set analog of Bayes' rule is computationally intractable in general settings \cite{goodman1997mathematics}. Intuitively, this follows because evaluation of fixed sets in the likelihoods for parametric densities of random sets may involve a term for each possible association of input points to parameters, resulting in exponential scaling with respect to the number of parameters. 
%
To address this, we model random persistence diagrams as Poisson point processes. The defining feature of these point processes is that they are solely characterized by a single parameter known as the intensity. Utilizing the theory of marked point processes, we obtain a method for computing posterior intensities that does not require us to consider explicit maps between input diagrams and underlying parameters, alleviating the computational burden associated with deriving the posterior intensity from Bayes' rule alone.  
 
 
In particular, for a given collection of observed persistence diagrams, we consider the underlying stochastic phenomena generating persistence diagrams to be Poisson point processes with prior uncertainty captured in presupposed intensities. In applications, one may select an informative prior by choosing an intensity based on expert opinion, or alternatively choose an uninformative prior intensity when information is not available.  The likelihood functions in our model represent the level of belief that observed diagrams are representative of the entire population. We build this analog using the theory of marked Poisson point processes \cite{daley}. A central idea of this paper is to utilize the topological summaries of point clouds in place of the actual point clouds. This provides a powerful tool with applications in wide ranging fields. The application considered in this paper is the classification of the crystal structure of materials, which allows scientists to predict the properties of a crystalline material. Our goal is to view point clouds through their topological descriptors as this can reveal essential shape peculiarities latent in the point clouds. Our Bayesian method adopts a substitution likelihood technique by Jeffreys in \cite{Jeffreys1961} instead of considering the full likelihood for the point cloud. A similar sort of discussion was considered in \cite{Mileyko2011} for defining conditional probability on persistence diagrams.
 
 Another key contribution of this paper is the derivation of a \emph{closed form} of the posterior intensity, which relies on conjugate families of Gaussian mixtures. An advantage of this Gaussian mixture representation is that it allows us to perform Bayesian inference in an efficient and reliable manner. Indeed, this model can be viewed as an analog of the ubiquitous example in standard Bayesian inference where a Gaussian prior and likelihood yield a Gaussian posterior. We present a detailed example of our closed form implementation to demonstrate computational tractability and showcase its applicability by using it to build a Bayes factor classification algorithm; we test the latter in a classification problem for materials science data. 
 
 The contributions of this work are:
 \begin{enumerate}
     \item  Theorem \ref{thm:bayes}, which provides the  Bayesian framework for computing the posterior distribution of persistence diagrams.  
     \item  Proposition \ref{prop:post}, which yields a conjugate family of priors based on a Gaussian mixture for the proposed Bayesian framework. 
     \item A classification scheme using Bayes factors considering the posteriors of persistence diagrams and its application to a materials science problem.
 \end{enumerate}
 
 This paper is organized as follows.  Section \ref{sec:background} provides a brief overview of persistence diagrams and general point processes. 
 Our methods are presented in Section \ref{sec:methods}. In particular, Subsection \ref{sec:bayesian model} establishes the Bayesian framework for persistence diagrams, while Subsection \ref{subsec:gm closed form computation} contains the derivation of a closed form for a posterior distribution based on a Gaussian mixture model. A classification algorithm with Bayes factors is discussed in Section \ref{sec:classification}. To assess the capability of our algorithm, we investigate its performance on materials data in Subsection \ref{subsec:real data}. Finally, we end with discussions and conclusions in Section \ref{sec: conclusion}.

\section{Background}\label{sec:background}
	We begin by discussing preliminary definitions essential for building our model. In Subsection \ref{subsec:persistence diagram}, we briefly review simplicial complexes and provide a formal definition for persistence diagrams (PDs). Pertinent definitions and theorems from point processes (PPs) are discussed in Subsection \ref{subsec:poisson pp} .

\subsection{Persistence Diagrams} \label{subsec:persistence diagram}
We start by discussing simplices and simplicial complexes, intermediary structures for constructing PDs.


\begin{definition}
	A $\d$-dimensional collection of data $\{v_{0},\dots,v_{n}\} \subset \R^{\d} \setminus \{0\} $ is said to be geometrically independent if for any set $t_{i} \in \R$ with $\sum_{i = 0}^{n} t_{i} = 0$, the equation
	$\sum_{i=0}^{n}t_{i}v_{i} = 0$
	implies that
	$
	t_{i} = 0$  for all $i\in \{ 0,\dots,n \}.
	$
\end{definition}

\begin{definition}
	\label{def:simplex}
	A $k-$simplex, is a collection of $k+1$ geometrically independent elements along with their convex hull:
	$[v_{0},\dots,v_{k}] = \Big\{\sum_{i = 0}^{k}\alpha_{i}v_{i} | \sum_{i = 0}^{k}\alpha_{i} = 1\Big\}$.
We say that the vertices $v_{0},\dots,v_{n}$ span the $k-$dimensional simplex, $[v_{0},\dots,v_{k}]$.
The faces of a $k-$simplex $[v_{0}, \dots, v_{k}]$, are the $(k-1)-$simplices spanned by subsets of $\{v_{0},\dots, v_{k}\}$. 
\end{definition}


\begin{definition}
	\label{simpcomp}
	A simplicial complex $S$ is a collection of simplices satisfying two conditions: (i) if $\xi \in S$, then all faces of $\xi$ are also in $S$, and (ii) the intersection of two simplices in $S$ is either empty or contained in $S$. 
\end{definition}

 Given a point cloud $X$, our goal is to construct a sequence of simplicial complexes that reasonably approximates the underlying shape of the data. 
  We accomplish this by using the Vietoris-Rips filtration.

\begin{definition}
	\label{ripsfilt}
	
	Let $X=\{ x_{i} \}_{i=0}^{L}$ be a point cloud in $\R^{\mathbbm{d}}$ and $r>0$. The Vietoris-Rips complex of $X$ is defined to be the simplicial complex $\mathcal{V}_{r}(X)$ satisfying
	$[x_{i_{1}},\dots,x_{i_{l}}] \in \mathcal{V}_{r}(X)$ if and only if $\diam(x_{i_{1}},\dots,x_{i_{l}}) < r$.
	Given a nondecreasing sequence $\{r_{n}\} \in \R^{+} \cup \{0\}$  with $r_{0} = 0$, we denote its Vietoris-Rips filtration by
	$\{\mathcal{V}_{r_{n}}(X) \}_{n \in \N}$.
\end{definition}

 A persistence diagram $\D$ is a multiset of points in $\mathcal{W} := \W \times \{0,1,\dots,\d-1 \}$, where $\W := \{(b,d) \in \R^{2} | \,\, d \geq b \geq 0\}$ and each element $(b,d,k)$ represents a homological feature of dimension $k$ that appears at scale $b$ during a Vietoris-Rips filtration and disappears at scale $d$. Intuitively speaking, the feature $(b,d,k)$ is a $k-$dimensional hole lasting for duration $d-b$. Namely, features with $k=0$ correspond to connected components, $k=1$ to loops, and $k=2$ to voids. An example of a PD is shown in Figure \ref{fig:tilted_representation}.

\subsection{Poisson Point Processes} \label{subsec:poisson pp}
This section contains basic definitions and fundamental theorems from PPs, primarily Poisson PPs. Detailed treatments of Poisson PPs can be found in \cite{daley} and references therein. 
For the remainder of this section, we take $\mathbb{X}$ and $\mathcal{X}$ to be a Polish space and its Borel $\sigma$-algebra, respectively.
\begin{definition}\label{def:finite_pp}
     A finite point process $\mathcal{P}$ is a pair $\left(\left\{p_{n}\right\},\left\{\Pro_{n}\right\}\right)$ where $\sum_{n=0}^{\infty} p_{n} = 1$ and $\Pro_{n}$ is a symmetric probability measure on $\mathcal{X}^{n}$, where $\mathcal{X}^{0}$ is understood to be the trivial $\sigma$-algebra.
\end{definition}

The sequence $\{p_n\}$ defines a cardinality distribution and the measures $\{\Pro_{n}\}$ give spatial distributions of vectors $(x_1,\dots,x_n)$ for fixed $n$. Definition \ref{def:finite_pp} naturally prescribes a method for sampling a finite PP: (i) determine the number of points $n$ by drawing from $\{p_n\}$ then, (ii) spatially distribute $(x_1,\dots,x_n)$ according to a draw from $\Pro_{n}$. As PPs model random collections of elements in $\{x_1,\dots,x_n\} \subset \mathbb{X}$ whose order is irrelevant, any sensible construction relying on random vectors should assign equal weight to all permutations of $(x_1,\dots,x_n)$. This is ensured by the symmetry requirement in Definition \ref{def:finite_pp}. We abuse notation and write $\mathcal{P}$ for samples from $\mathcal{P}$ as well as their set representations.  
It proves useful to describe finite PPs by a set of measures that synthesize $p_n$ and $\Pro_n$ to simultaneously package cardinality and spatial distributions. %
\begin{definition} \label{def:janossy}
Let $\left(\left\{p_{n}\right\},\left\{\Pro_{n}\right\}\right)$ be a finite PP. The Janossy measures $\left\{\mathbb{J}_{n}\right\}$ are defined as the set of measures satisfying $\mathbb{J}_{n} = n!p_{n}\Pro_{n}, \hspace{2mm} \text{for all} \hspace{2mm} n \in \mathbb{N}.$
\end{definition}

Given a collection of disjoint rectangles $A_1,\dots,A_n \subset \mathbb{X}$, the value $\mathbb{J}_{n}(A_{1}\times \dots \times A_{n})$ is the probability of observing exactly one element in each of $A_{1},\dots,A_{n}$ and none in the complement of their union. For applications, we are primarily interested in Janossy measures $\mathbb{J}_n$ that admit densities $j_n$ with respect to a reference measure on $\mathbb{X}$. We are now ready to describe the class of finite PPs that model PDs.
\begin{definition}\label{def:poisson}
Let $\Lambda$ be a finite measure on $\mathbb{X}$ and $\mu := \Lambda(\mathbb{X})$. 7The finite point process $\Pi$ is Poisson if, for all $n \in \mathbb{N}$ and measurable rectangles $A_{1} \times \dots \times A_{n} \in \mathcal{X}^{n}$,
$p_n =  e^{-\mu}\frac{\mu^{n}}{n!},$ and $\Pro_{n}(A_{1}\times \dots \times A_{n} ) = \prod_{i=1}^{n}\left(\frac{\Lambda(A_{i})}{\mu}\right).$We call $\Lambda$ an intensity measure.
\end{definition}

Equivalently, a Poisson PP is a finite PP with Janossy measures $\mathbb{J}_{n}(A_{1}\times \dots \times A_{n}) = e^{-\mu}\prod_{i=1}^{n} \Lambda(A_{i}).$
The intensity measure in Definition \ref{def:poisson} admits a density, $\lambda$, with respect to some reference measure on $\mathbb{X}$. Notice that for all $A \in \mathcal{X}$,  
$\E(|\Pi \cap A|) = \sum_{n=0}^{\infty}p_n\E_{\Pro_n}\left(\sum_{k=0}^{n}k\binom{n}{k}\1_{A^{k}\times(A^{c})^{n-k}}\right)$. Elementary calculations then show $\E(|\Pi \cap A|) = \Lambda(A)$. 
Thus, we interpret the intensity measure of a region $A$, $\Lambda(A)$ as the expected number of elements in $\Pi$ that land in $A$. The intensity measure serves as an analog to the first order moment for a random variable.

The next two definitions involve a joint PP wherein points from one space parameterize distributions for the points living in another. Consequently, we introduce another Polish space $\mathbb{M}$ along with its Borel $\sigma$-algebra $\mathcal{M}$ to serve as the mark space in a marked Poisson PP. These model scenarios in which points drawn from a Poisson PP provide a data likelihood model for Bayesian inference with PPs. 
%
\begin{definition}
\label{def:stochastic_kernel}
Suppose $\ell:\mathbb{X} \times \mathbb{M} \rightarrow \R^{+}\cup \{0\}$ is a function satisfying: 1) for all $x \in \mathbb{X}$, $\ell(x,\bullet)$ is a probability measure on $\mathbb{M}$, and 2) for all $B \in \mathcal{M}$, $\ell(\bullet,B)$ is a measurable function on $\mathbb{X}$. Then, $\ell$ is a stochastic kernel from $\mathbb{X}$ to $\mathbb{M}$. 
\end{definition}
\begin{definition}
\label{def:marked_poisson_process}
 A marked Poisson point process $\Pi_{M}$ is a finite point process on $\mathbb{X} \times \mathbb{M}$ such that: (i) $\left(\left\{p_{n}\right\},\left\{\Pro_{n}(\bullet \times \mathbb{M})\right\}\right)$ is a Poisson PP on $\mathbb{X}$, and (ii) for all $(x_{1},\dots,x_{n}) \in \mathbb{X}^{n}$, measurable rectangles $B_{1} \times \dots \times B_{n} \in \mathcal{M}^{n}$ , $\Pro_{n}((x_{1},\dots,x_{n}) \times B_{1} \times \dots \times B_{n}) = \frac{1}{n!}\sum_{\pi \in \mathcal{S}_{n}}\prod_{i=1}^{n}\ell(x_{\pi(i)},B_{i})$, where $\mathcal{S}_n$ is the set of all permutations of $(1,\dots,n)$ and $\ell$ is a stochastic kernel.
\end{definition}
Given a set of observed marks $M =\ \{y_{1},\dots,y_{m}\}$ it can be shown \cite{filters_for_spp} that the Janossy densities for the PP induced by $\Pi_{M}$ on $\mathbb{X}$ given M are

\begin{equation} \label{eqn:conditional_janossy}
    j_{n|M}(x_1,\dots,x_n) = \begin{cases} \sum_{\pi \in \mathcal{S}_{n}}\prod_{i=1}^{n}p(x_{i}|y_{\pi(i)}), \,\, n = m,\\
    0,  \hspace{1.32in}  \text{otherwise},
                            \end{cases}
\end{equation}
where $p$ is the stochastic kernel for $\Pi_{M}$ evaluated in $\mathbb{X}$ for a fixed value of $y \in \mathbb{M}$.

\noindent The following theorems allow us to construct new Poisson PPs from existing ones. Their proofs can be found in \cite{poisson_processes}.
\begin{theorem}[The Superposition Theorem]
\label{thm:superposition}
Let $\{\Pi_{n}\}_{n\in \N}$ be a collection of independent Poisson PPs each having intensity measure $\Lambda_{n}$. Then their superposition $\Pi$ given by $\Pi := \bigcup_{n \in \N} \Pi_{n}$
is a Poisson PP with intensity measure $\Lambda = \sum_{n \in \N} \Lambda_{n}$.
\end{theorem}
\begin{theorem}[The Mapping Theorem]
\label{thm:mapping} 
Let $\Pi$ be a Poisson PP on $\mathbb{X}$ with $\sigma$-finite intensity measure $\Lambda$ and let $(\mathbb{T},\mathcal{T})$ be a $\sigma$-algebra. Suppose $f:\mathbb{X} \rightarrow \mathbb{T}$ is a measurable function. Write $\Lambda^{*}$ for the induced measure on $T$ given by $\Lambda^{*}(B) := \Lambda(f^{-1}(B))$ 
for all $B \in \mathcal{T}$. If $\Lambda^{*}$ has no atoms, then $f \circ \Pi$ is a Poisson PP on $T$ with intensity measure $\Lambda^{*}$. 
\end{theorem}
\begin{theorem}[The Marking Theorem]
\label{thm:marking}
The marked Poisson PP in Definition \ref{def:marked_poisson_process} has the intensity measure given by $\Lambda_{M}(C) = \iint_{C} \Lambda(dx)\ell(x,dm)$, where $\Lambda$ is the intensity measure for the Poisson PP that $\Pi_{M}$ induces on $\mathbb{X}$, and $\ell$ is a stochastic kernel.
\end{theorem}

The final tool we need is the probability generating functional as it enables us to recover intensity measures using a notion of differentiation. The probability generating functional can be interpreted as the PP analog of the probability generating function. 
\begin{definition} \label{def:pgfl}
Let $\mathcal{P}$ be a finite PP on a Polish space $\mathbb{X}$. Denote by $\mathcal{B}(\mathbb{C})$ the set of all functions $h:\mathbb{X} \rightarrow \mathbb{C}$ with $||h||_{\infty} < 1$. The probability generating functional of $\mathcal{P}$ denoted $G: \mathcal{B}(\mathbb{C}) \rightarrow \mathbb{R}$ is given by
\begin{equation}\label{eqn:pgfl}
    G(h) = J_{0} + \sum_{n=1}^{\infty}\frac{1}{n!}\int_{\mathbb{X}^{n}}\left(\prod_{j=1}^{n} h(x_{j})\right)\mathbb{J}_{n}(dx_{1}\dots dx_{n})
\end{equation}
\end{definition}
\begin{definition} \label{def:functional_deriv}
Let $G$ be the probability generating functional given in Equation \eqref{eqn:pgfl}. The functional derivative of $G$ in the direction of $\eta$ evaluated at $h$, when it exists, is given by $G'(h;\eta) = \lim_{\epsilon \rightarrow 0} \frac{G(h + \epsilon\eta) - G(h)}{\epsilon}$. 
\end{definition}

It can be shown that the functional derivative satisfies the familiar product rule \cite{mahler2007statistical}. As is proved in \cite{theory_of_pop_processes}, the intensity measure $\Lambda$ of the Poisson PP in Definition \ref{def:poisson} can be obtained by differentiating $G$, i.e., $\Lambda(A) = G'(1;\1_{A})$, where $\1_{A}$ is the indicator function for any $A \in \mathcal{X}$. Generally speaking, one obtains the intensity measure for a general point process through $\Lambda(A) = \lim_{h \rightarrow 1} G'(h;\1_{A})$, but the preceding identity suffices for our purposes since we only consider point processes for which Equation \eqref{eqn:pgfl} is defined for all bounded $h$.

\begin{corollary}\label{cor:conditional_intensity_mpp}
The intensity function for the PP whose Janossy densities  are listed in Equation \eqref{eqn:conditional_janossy} is $ \sum_{i=1}^{m}p(x|y_{i})$.
\end{corollary}
\begin{proof}
This directly follows from writing the probability generating functional for the PP in question using its Janossy densities then applying $\Lambda(A) = G'(1;\1_{A})$. By Definition \ref{def:pgfl}, linearity of the integral, and Fubini's theorem, we have $G(h) = \prod_{i=1}^{m}\Big(\int_{\mathbb{X}}h(x)p(x|y_{i})\infinitesmal x\Big) = \prod_{i=1}^{m}G_{i}(h)$ where $G_{i}(h)$ is the probability generating functional for the PP with Janossy densities $j_1(x)=p(x|y_{i})$ and $j_n = 0$ for $n \neq 1$. One arrives at the desired result by applying the product rule for functional derivatives and the intensity retrieval property of probability generating functionals. 
\end{proof}
%

\section{Bayesian Inference}\label{sec:methods}
In this section, we construct a framework for Bayesian inference with PDs by modeling them as Poisson PPs. First, we derive a closed form for the posterior intensity given a PD drawn from a finite PP, and then we present a family of conjugate priors followed by an example.  

\subsection{Model} \label{sec:bayesian model}
\begin{figure}

\centering
\subfloat[]{\includegraphics[width=1.6in,height=1.5in]{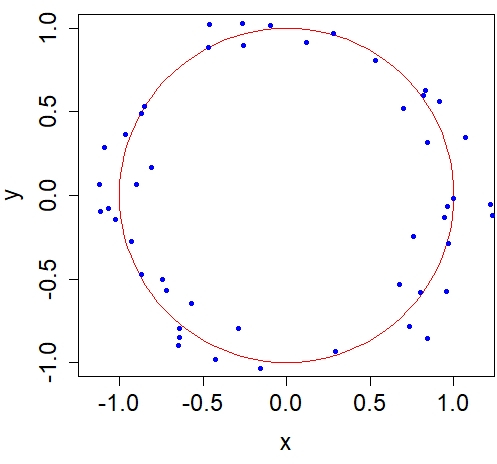}}
	\centering
\subfloat[]{\includegraphics[width=1.6in,height=1.5in]{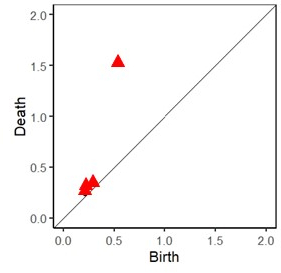}}
 	\centering
\subfloat[]{\includegraphics[width=1.6in,height=1.5in]{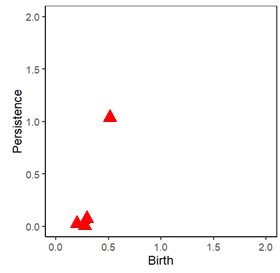}}
    \caption{\textit{ (a) An example of a dataset; (b) Its persistence diagram; (c) Its tilted representation.}}
 \label{fig:tilted_representation}
 
 	\vspace{-0.3in}
\end{figure}

 Given a persistence diagram $\mathcal{D}$, the map $T:\W \rightarrow T(\W)$ given by $T(b,d)=(b,d-b)$ defines tilted representation of $\D$ as $T(\mathcal{D}) = \cup_{(b,d,k) \in \mathcal{D}}(T(b,d),k)$; see Figure \ref{fig:tilted_representation}. In the sequel, we assume all PDs are given in their tilted representations and, unless otherwise noted, abuse notation by writing $\W$ and $\mathcal{D}$ for $T(\W)$ and $T(\mathcal{D})$, respectively. We also fix the homological dimension of features in a PD by defining $\mathcal{D}^{k} := \{(b,d) \in \W \hspace{2mm} | \hspace{2mm} (b,d,k) \in \mathcal{D}\}$.

 According to Bayes' theorem, posterior density is proportional to the product of a likelihood function and a prior. 
To adopt Bayesian framework to PDs, we need to define two models.
In particular, our Bayesian framework views a random PD as a Poisson PP equipped with a prior intensity while observed PDs $\D_Y$ are considered to be marks from a marked Poisson PP. This enables modification of the prior intensity by incorporating observed PDs, yielding a posterior intensity based on data. Some parallels between our Bayesian framework and that for random variables (RVs) are illustrated in Table \ref{tbl:bayes}.
 \begin{table}[tbhp]
 \captionsetup{name=TABLE, justification=centering,labelsep=newline}
 \centering
  {
 \caption { \footnotesize \textit{The parallels between the Bayesian framework for RVs and its counterpart for random PDs.}}\label{tbl:bayes}
	\begin{tabular}{|c|c|c|}
		\hline 
		 & Bayesian Framework for RVs & Bayesian Framework for Random PDs\\
		\hline 
		\textit{Prior}  & Modeled by a prior density $f$ & Modeled by a Poisson 
		PP with prior intensity $\lambda$ \\ 
			\hline 
		\textit{Likelihood} & Depends on observed data  & Stochastic kernel that depends on observed PDs \\ 
			\hline 
			
		\textit{Posterior}  & Compute the posterior density                                   & A Poisson PP with posterior intensity\\
		\hline 
	\end{tabular} 
}
\end{table}
 Let $(\mathcal{D}_{X}^k,\mathcal{D}_{Y}^k) \in \W\times \W$ be a finite PP and consider the following: 


\begin{itemize}
\item[(M1)] For $k_1 \neq k_2$, $(\mathcal{D}_{X}^{k_1},\mathcal{D}_{Y}^{k_1})$ and $(\mathcal{D}_{X}^{k_2},\mathcal{D}_{Y}^{k_2})$ are independent.
\item[(M2)] For $k$ fixed, $\mathcal{D}_{X}^{k} = \mathcal{D}_{X_O}^{k}\cup \mathcal{D}_{X_V}^{k}$
and some $\alpha:\W \rightarrow [0,1]$, $\mathcal{D}_{X_O}^{k}$ and $\mathcal{D}_{X_V}^{k}$ are independent Poisson PPs having intensity functions $\alpha(x)\lambda_{\mathcal{D}_{X}^{k}}(x)$ and $\left(1-\alpha(x)\right)\lambda_{\mathcal{D}_{X}^{k}}(x)$, respectively.

\item[(M3)] For $k$ fixed, $\mathcal{D}_{Y}^{k} = \mathcal{D}_{Y_O}^{k}\cup \mathcal{D}_{Y_S}^{k}$ where
    \begin{itemize}
        \item[(i)] $(\mathcal{D}_{X_O}^{k},\mathcal{D}_{Y_O}^{k})$ is a marked Poisson PP with a stochastic kernel density $\ell(y|x)$.
        \item[(ii)] $\mathcal{D}_{Y_O}^{k}$ and $\mathcal{D}_{Y_S}^{k}$ are independent finite Poisson PPs where $\mathcal{D}_{Y_S}^{k}$ has intensity function $\lambda_{\mathcal{D}_{Y_S}^{k}}$.
    \end{itemize}
\end{itemize}

\begin{figure}[h!]
    \centering
    \subfloat[]{\includegraphics[scale=0.3]{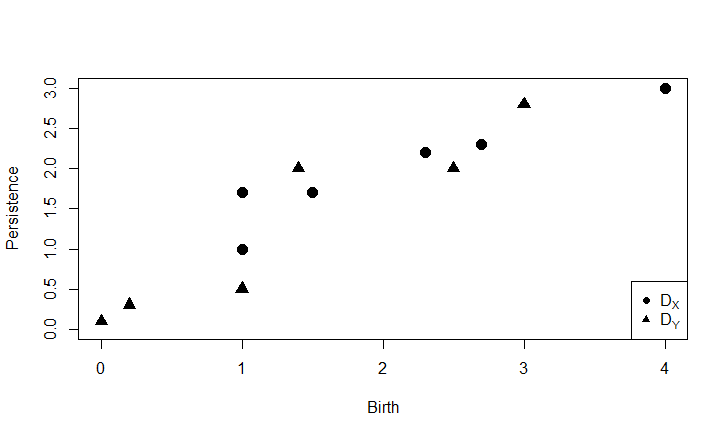}}
    \subfloat[]{\includegraphics[scale=0.3]{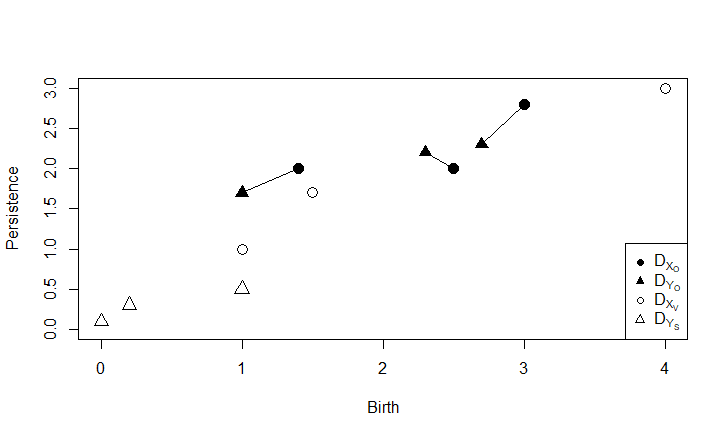}}
    \caption{\textit{(a) A sample from the prior point process $\mathcal{D}_X$ and an observed persistence diagram $\mathcal{D}_Y$. (b) Decomposition of $\mathcal{D}_X$ into $\mathcal{D}_{X_O}$,$\mathcal{D}_{X_V}$ and $\mathcal{D}_Y$ into $\mathcal{D}_{Y_O}$,$\mathcal{D}_{Y_S}$. The points in $\mathcal{D}_{X_V}$ have no relationship to those in $\mathcal{D}_{Y}$, while those in $\mathcal{D}_{X_O}$ only generate observed points in $\mathcal{D}_{Y_O}$. The remaining observed points in $\mathcal{D}_{Y_S}$ model unanticipated features that one may obtain due to uncertainty/noise.} }
    \label{fig:dxdy}
\end{figure}

Hereafter we abuse notation by writing $\mathcal{D}_{X}$ for $\mathcal{D}_{X}^{k}$. 
%
The modeling assumption (M1) allows us to develop results independently for each homological dimension $k$ then combine them using independence. In (M2), the random persistence diagram $\D_X$ modeled as a Poisson PP with prior intensity $\lambda_{\D_X}$. There are two cases we may encounter for any point $x$ from the prior intensity due to the nature of persistence diagrams. We assign a probability function $\alpha(x)$ to accommodate these two possibilities. Depending upon the noise level in data, any feature $x$ in $\mathcal{D}_X$ may not be represented in observations and this scenario happens with probability $1-\alpha(x)$ and we denote this case as $\D_{X_{V}}$ in (M2). Otherwise a point $x$ observed with a probability of $\alpha(x)$ and this scenario is presented as $\D_{X_{O}}$ in (M2).
Consequently, the intensities of $\D_{X_O}$ and $\D_{X_V}$ are proportional to the intensity $\lambda_{\D_X}$ weighted by $\alpha(x)$ and $1-\alpha(x)$ respectively and the total prior intensity for $\D_{X}$ is given by their sum. (M3) considers observed persistence diagram $\D_Y$ and decomposes it into two independent PDs, $\D_{Y_O}$ and $\D_{Y_S}$. $\D_{Y_O}$ is linked to $\D_{X_O}$ via a marked point process with likelihood $\ell(y|x)$ defined in Equation \eqref{eqn:conditional_janossy}, whereas the component $\D_{Y_S}$ includes any point $y$ that arises from noise or unanticipated geometry. See Figure \ref{fig:dxdy} for a graphical representation of these ideas.

	\vspace{-0.3in}


\vspace{-0.05in}
\begin{theorem}[Bayesian Theorem for Persistence Diagrams] \label{thm:bayes}
Let $\mathcal{D}_{X}$ be a persistence diagram modeled by a Poisson PP as in (M2). Suppose 
$\mathcal{D}_{X_O}$ and $\mathcal{D}_{X_V}$ have prior intensities $\alpha(x)\lambda_{\mathcal{D}_{X}}$ and $(1-\alpha(x))\lambda_{\mathcal{D}_{X}}$, respectively. Consider $D_{Y^1},\dots,D_{Y^m}$ independent samples from the point process that characterizes the persistence diagram $\mathcal{D}_{Y}$ of (M3) and denote $D_{Y^{1:m}} := \cup_{i=1}^{m}D_{Y^i}$ where $D_{Y^i}=D_{Y_O^i} \cup D_{Y_S^i}$ for all $i=1, \cdots, m$.  Moreover, $\ell(y|x)$ is the likelihood associated for the stochastic kernel between $\mathcal{D}_{X_O}$ and $\mathcal{D}_{Y_O}$, and $\lambda_{\D_{Y_S}}$ is the intensity of $\D_{Y_S}$ as defined in (M3). Then, the posterior intensity of $\mathcal{D}_{X}$ given $D_{Y^{1:m}}$ is
\begin{equation} \label{post intensity}
    \lambda_{\D_X|D_{Y^{1:m}}}(x) = \left(1-\alpha(x)\right)\lambda_{\mathcal{D}_X}(x)+ \frac{1}{m}\alpha(x) \sum_{i=1}^m\sum_{y \in D_{Y^i}}\frac{\ell(y|x)\lambda_{\mathcal{D}_X}(x)}{\lambda_{\mathcal{D}_{Y_S}}(y)+\int_{\W}\ell(y|u)\alpha(u)\lambda_{\mathcal{D}_X}(u) du} \,\,\, \text{a.s.}
\end{equation}
\end{theorem}

The proof of Theorem \ref{thm:bayes} can be found in the appendix. One important point about the above theorem is that, instead of relying on a likelihood function for the point cloud data, our Bayesian model considers the likleihood for the persistence diagram generated by the observed point cloud data at hand. 
This 
is analogous to the idea of substitution likelihood by Jeffreys in \cite{Jeffreys1961}.
 
%
%
\subsection{A Conjugate Family of Prior Intensities: Gaussian Mixtures}\label{subsec:gm closed form computation}
This section focuses on constructing a family of conjugate prior intensities, i.e., a collection of priors that yield posterior intensities of the same form when used in Equation \eqref{post intensity}.
Exploiting Theorem \ref{thm:bayes} with Gaussian mixture prior intensities, we obtain Gaussian mixture posterior intensities. 
As PDs are stochastic point processes on the space $\W$, not $\R^2$, we consider a restricted Gaussian density restricted to $\W$. Namely, for a Gaussian density on $\R^2$, $\mathcal{N}(z;\upsilon,\sigma I)$, with mean $\upsilon$ and covariance matrix $\sigma I$ , we restrict the Gaussian density on $\W$ as 
	%
	{\begin{equation}\label{restricted gaussian}
	\mathcal{N}^{*}(z;\upsilon,\sigma I) := \mathcal{N}(z;\upsilon,\sigma I)\1_{\W}(z),
	\end{equation}}
where $\1_{\W}$ is the indicator function of the wedge $\W$. `


 Consider a random persistence diagram $\D_X$ as in (M2) and a collection of  observed PDs $\{D_{Y^1}, \cdots, D_{Y^m}\}$ that are independent samples from Poisson PP characterizing the PD $\mathcal{D}_Y$ in (M3). We denote $D_{Y^{1:m}} :=\cup_{i=1}^m D_{Y^i}$. 
Below we specialize (M2) and (M3) so that applying Theorem \ref{thm:bayes} to a mixed Gaussian prior intensity yields a mixed Gaussian posterior:
\begin{enumerate}
\item [(M2$'$)] $\D_X=\D_{X_O} \cup \D_{X_V}$, where $\D_{X_O}$ and $\D_{X_V}$ are independent Poisson PPs with intensities $\alpha\lambda_{\mathcal{D}_X}(x)$ and  $(1-\alpha)\lambda_{\mathcal{D}_X}(x)$, respectively, with

\vspace{-0.2in}
    \begin{equation} \label{eqn:intensity of Dx}
    \lambda_{\mathcal{D}_X}(x) = \sum_{j = 1}^{N}c^{\mathcal{D}_X}_{j}\mathcal{N}^{*}(x;\mu^{\mathcal{D}_X}_{j},\sigma^{\mathcal{D}_X}_{j}I), 
    \vspace{-0.1in}
    \end{equation}
    where $N$ is the number of mixture components. 
    \item [(M3$'$)]  $\mathcal{D}_Y=\mathcal{D}_{Y_O} \cup \mathcal{D}_{Y_S}$ where
    \begin{enumerate}
    \item[(i)] the marked Poisson PP $(\D_{X_O},\D_{Y_O})$ has density $\ell({y|x})$ given by 
    \begin{equation} \label{eqn:stachastic kernel gaussian}
      \ell(y|x) = \mathcal{N}^{*}(y;x,\sigma^{\mathcal{D}_{Y_O}}I).
       \end{equation}
    \item[(ii)] $\mathcal{D}_{Y_O}$ and $\mathcal{D}_{Y_S}$ are independent finite Poisson PPs and $\mathcal{D}_{Y_S}$ has intensity function given below.
    
       \begin{equation}\label{eqn:intensity of Dys}
     \lambda_{\mathcal{D}_{Y_S}}(y) = \sum_{k = 1}^{M}c^{\mathcal{D}_{Y_S}}_{k}\mathcal{N}^{*}(y;\mu^{\mathcal{D}_{Y_S}}_{k},\sigma^{\mathcal{D}_{Y_S}}_{k}I),
     \end{equation}
      where $M$ is the number of mixture components. 
    \end{enumerate}

\end{enumerate}
\begin{proposition}	\label{prop:post}
Suppose that the assumptions (M1),(M2$'$), and (M3$'$) hold; then, the posterior intensity of Equation \eqref{post intensity} in Theorem \ref{thm:bayes} is a Gaussian mixture of the form 
\begin{equation}
	\label{eqn:mg_posterior}
	\lambda_{\mathcal{D}_X|D_{Y^{1:m}}}(x) = (1-\alpha)\lambda_{\mathcal{D}_X}(x)+\frac{\alpha}{m}  \sum_{i=1}^m\sum_{y \in \D_{Y^i}}\sum_{j=1}^{N} C_{j}^{y}\mathcal{N}^*(x;\mu_{j}^{y},\sigma_{j}^{y}I),
	\end{equation}
	
	\vspace{-0.2in}
	 \begin{align}
	  \text{where}\,\,\,\,\,\,\,\,\,   C_{j}^{y} &= \frac{w_{j}^{y}}{\lambda_{\mathcal{D}_{Y_S}}(y)+\alpha\sum_{j=1}^{N}w_{j}^{y}Q_{j}^{y}}; \,\,Q_{j}^{y} = \int_{\W}\mathcal{N}(u;\mu_{j}^{y},\sigma_{j}^{y}I) du; \nonumber\\
	     w_{j}^{y} &=  c_{j}^{\mathcal{D}_X}\mathcal{N}(y;\mu_{j}^{\mathcal{D}_X},(\sigma^{\mathcal{D}_{Y_O}}+\sigma_{j}^{\mathcal{D}_X})I);\nonumber \\
	     \text{and} \,\,\,\,\,\,\,\,\,\mu_{j}^{y} &= \frac{\sigma_{j}^{\mathcal{D}_X} y+\sigma^{\mathcal{D}_{Y_O}}\mu_{j}^{\mathcal{D}_X}}{\sigma_{j}^{\mathcal{D}_X}+\sigma^{\mathcal{D}_{Y_O}}}; 
 	\sigma_{j}^{y}= \frac{\sigma^{\mathcal{D}_{Y_O}}\,\sigma_{j}^{\mathcal{D}_X}}{\sigma_{j}^{\mathcal{D}_X}+\sigma^{\mathcal{D}_{Y_O}}}.\nonumber
	 \end{align}
\end{proposition}

The proof of Proposition \ref{prop:post} follows from well known results about products of Gaussian densities given below; for more details, the reader may refer to \cite{Marouls2015} and references therein.
\begin{lemma} \label{lemma:prop of gaussian dist}
For $p \times p$ matrices $H, R, P$, with $R$ and $P$ positive definite ,and  a $p \times 1$ vector $s$,

\noindent $\mathcal{N}(y;Hx,R)\,\mathcal{N}(x;s,P)=q(y)\,\mathcal{N}(x;\hat{s},\hat{P})$, where $q(y)=\mathcal{N}(y;Hs,R+HPH^T), \,\, \hat{s}=s+K(y-Hs), \,\, \hat{P}=(I -KH)P$ and $K=PH^T(HPH^T+R)^{-1}$.
 \end{lemma}
\begin{proof}[Proof of Proposition  \ref{prop:post}]
Using Lemma \ref{lemma:prop of gaussian dist}, we first derive $\ell(y|x) \,\lambda_{\mathcal{D}_X}(x)$ by observing that, in our model, $H=I, R=\sigma^{\mathcal{D}_{Y_O}}I, s={\mu}_{j}^{\mathcal{D}_X}$ and  $P=\sigma_{j}^{\mathcal{D}_X}I$. By typical matrix operations we obtain, $K=\frac{\sigma_{j}^{\mathcal{D}_X}}{\sigma_{j}^{\mathcal{D}_X}+\sigma^{\mathcal{D}_{Y_O}}}, \hat{s}=\frac{\sigma_{j}^{\mathcal{D}_X} y+\sigma^{\mathcal{D}_{Y_O}}\mu_{j}^{\mathcal{D}_X}}{\sigma_{j}^{X}+\sigma^{\mathcal{D}_{Y_O}}}$, and $\hat{P}=\frac{\sigma^{\mathcal{D}_{Y_O}}\sigma_{j}^{\mathcal{D}_X}}{\sigma_{j}^{\mathcal{D}_X}+\sigma^{\mathcal{D}_{Y_O}}}$. Hence the numerator and denominator of the second term in Equation \eqref{post intensity}, $\sum_{j=1}^{N}c_{j}^{\mathcal{D}_X}\mathcal{N}(y;\mu_{j}^{\mathcal{D}_X},(\sigma^{\mathcal{D}_{Y_O}}+\sigma_{j}^{\mathcal{D}_X})I) \,\mathcal{N}^{*}(x;\mu_{j}^{y},\sigma_{j}^{y}I)$, and 
$\lambda_{\mathcal{D}_{Y_S}}(y)+\alpha\sum_{j=1}^{N}c_{j}^{\mathcal{D}_X}\mathcal{N}(y;\mu_{j}^{\mathcal{D}_X},(\sigma^{\mathcal{D}_{Y_O}}+\sigma_{j}^{\mathcal{D}_X})I) \int_{\W}\,\mathcal{N}(u;mu_{j}^{y},\sigma_{j}^{y}I)du,$ respectively, yield
\vspace{0.1in}
 \[\sum_{j=1}^{N} \Big[\frac{c_{j}^{\mathcal{D}_X}\mathcal{N}(y;\mu_{j}^{\mathcal{D}_X},(\sigma^{\mathcal{D}_{Y_O}}+\sigma_{j}^{\mathcal{D}_X})I)}{\lambda_{\mathcal{D}_{Y_S}}(y)+\alpha\sum_{j=1}^{N}c_{j}^{\mathcal{D}_X}\mathcal{N}(y;\mu_{j}^{\mathcal{D}_X},(\sigma^{\mathcal{D}_{Y_O}}+\sigma_{j}^{\mathcal{D}_X})I)\int_{\W}\mathcal{N}(u;\mu_{j}^{x|y},\sigma_{j}^{x|y}I)du}\Big]\mathcal{N}^*(x;\mu_{j}^{x|y},\sigma_{j}^{x|y}I),\]

\vspace{0.1in} 
\noindent where the bracketed expression is the definition of $C_j^{y}$. 
\end{proof}

 \subsubsection{Example} \label{example gaussian}

Here, we present a detailed example of computing the posterior intensity according to Equation \eqref{eqn:mg_posterior} for a range of parametric choices. Reproducing these results, the interested reader may download our R-package   \href{https://github.com/maroulaslab/BayesTDA}{BayesTDA}. We consider circular point clouds often associated with periodicity in signals \cite{Marchese2018} and focus on estimating homological features with $k=1$ as they correspond to 1-dimensional holes, which describe the prominent topological feature of a circle. Precisely our goals are to: (i) illustrate posterior intensities and draw analogies to standard Bayesian inference; (ii) determine the relative contributions of the prior and observed data to the posterior; and (iii) perform sensitivity analysis.
 \begin{table}[h!]
\vspace{-0.1in}
\captionsetup{justification=centering,labelsep=newline}
 \caption{\footnotesize \textit{List of Gaussian mixture parameters of the prior intensities in Equation \eqref{eqn:intensity of Dx}. The means $\mu_i^{\D_X}$ are $2 \times 1$ vectors and the rest are scalars}}.\label{tabel:prior parameters}
\begin{center}
\begin{tabular}{|c|c|c|c|}
\hline
 & $\mu_i^{\D_X}$           & $\sigma_i^{\D_X}$  & $c_i^{\D_X}$\\ \hline
Informative Prior             & $(0.5,1.2)$ & $0.01$  & $1$    \\ \hline
Weakly informative Prior              & $(0.5,1.2)$ & $0.2$   & $1$    \\
\hline
Unimodal Uninformative Prior              & $(1,1)$ & $1$  & $1$     \\
\hline
\raisebox{1.5ex}{Bimodal Uninformative Prior} &\shortstack{$(0.5,0.5)$ \\$(1.5,1.5)$} &\shortstack{$0.2$ \\ $0.2$ }  &\shortstack{$1$ \\ $2$} \\
\hline
\end{tabular}
\end{center}
\end{table}

\begin{figure}[h!]
\centering
\stackunder[1pt]{\includegraphics[width=1.6in,height=1.5in]{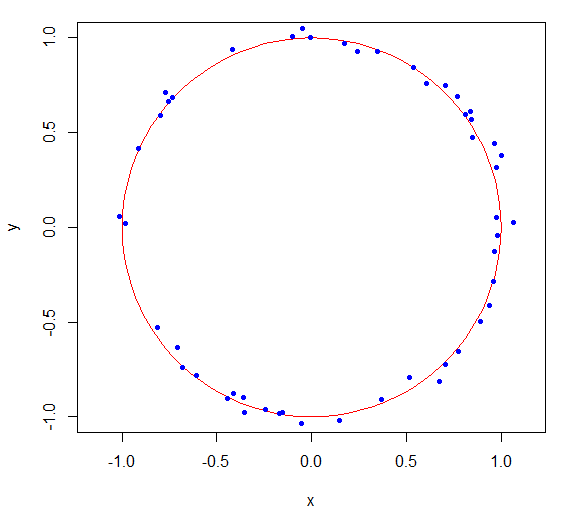}}{\footnotesize{\emph{Case-I}}}
\stackunder[1pt]{\includegraphics[width=1.6in,height=1.5in]{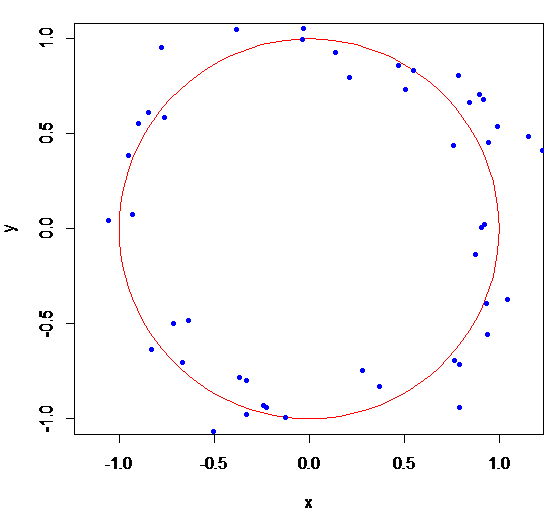}}{\footnotesize{\emph{Case-II}}}
\stackunder[1pt]{\includegraphics[width=1.6in,height=1.5in]{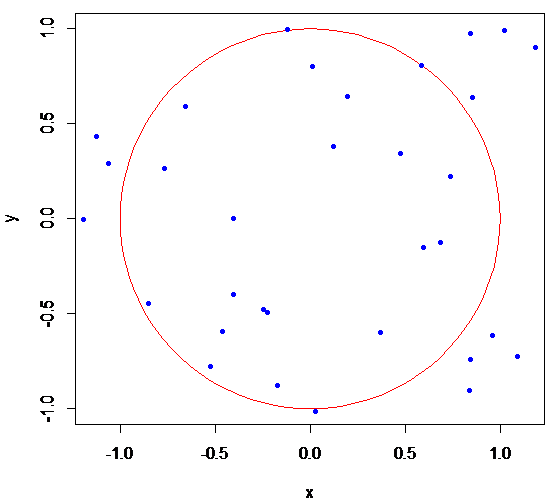}}{\footnotesize{\emph{Case-III}}}
\caption{\footnotesize{\textit{The observed datasets generated for Case-I, Case-II, and Case-III by sampling the unit circle and perturbing with Gaussian noise having variances $0.001I_2$, $0.01I_2$, and $0.1I_2$, respectively.}} }
 \label{fig:circle_data}
 \vspace{-0.3in}
 \end{figure}

\begin{figure} [h!]
      \centering
      \subfloat[]{\includegraphics[width=1.3in,height=1.4in]{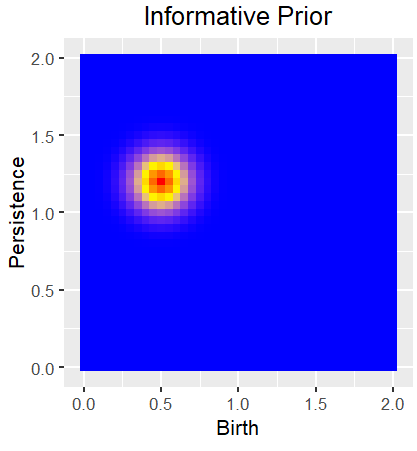}}
    \subfloat[]{\includegraphics[width=1.3in,height=1.4in]{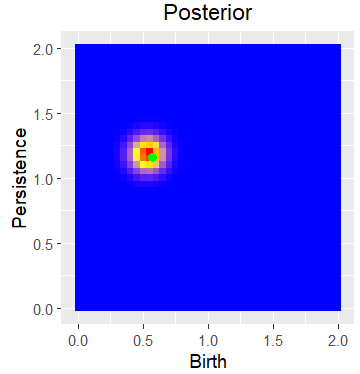}}
     \subfloat[]{\includegraphics[width=1.5in,height=1.4in]{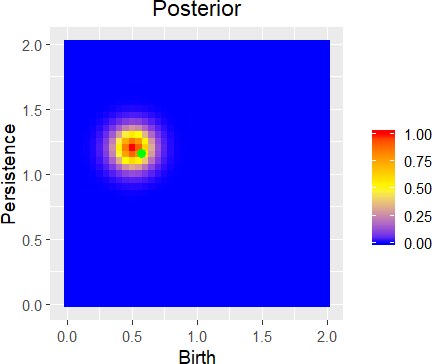}}\\
     \subfloat[]{\includegraphics[width=1.3in,height=1.4in]{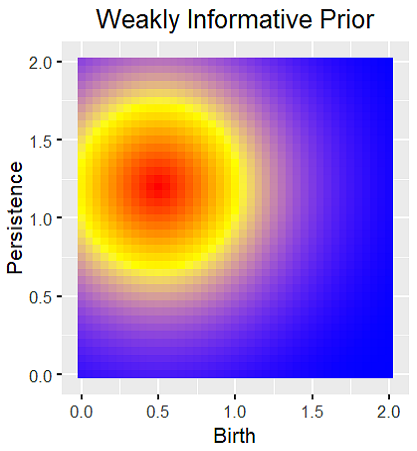}}
    \subfloat[]{\includegraphics[width=1.3in,height=1.4in]{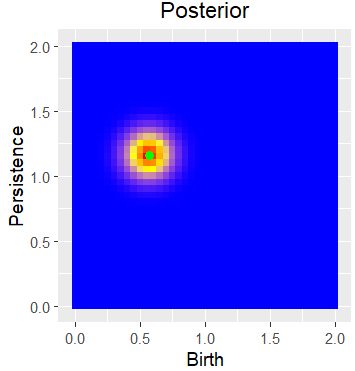}}
     \subfloat[]{\includegraphics[width=1.5in,height=1.4in]{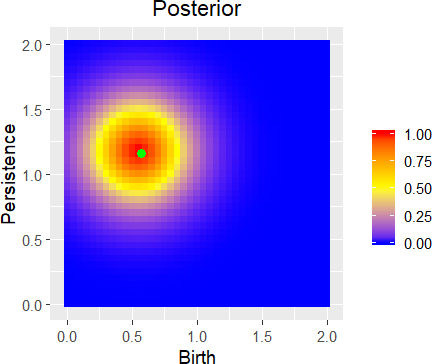}}\\
     \subfloat[]{\includegraphics[width=1.3in,height=1.4in]{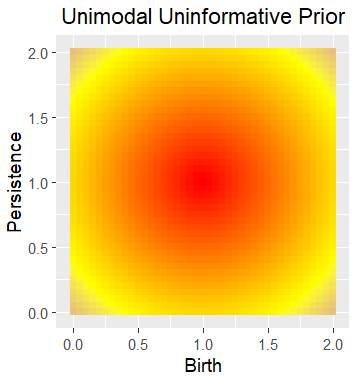}}
     \subfloat[]{\includegraphics[width=1.3in,height=1.4in]{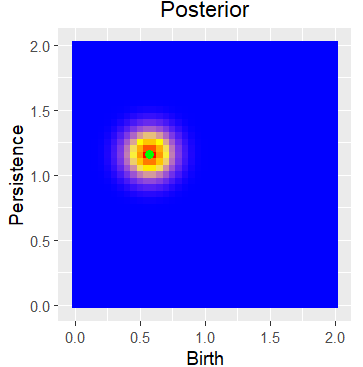}}
     \subfloat[]{\includegraphics[width=1.5in,height=1.4in]{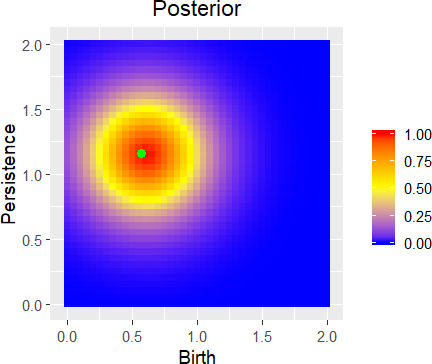}}\\
      \subfloat[]{\includegraphics[width=1.3in,height=1.4in]{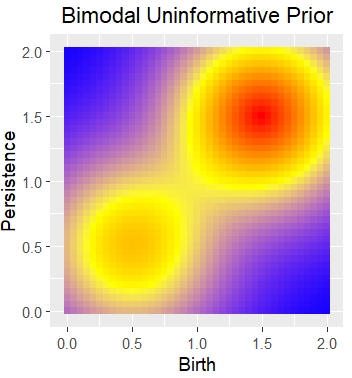}}
     \subfloat[]{\includegraphics[width=1.3in,height=1.4in]{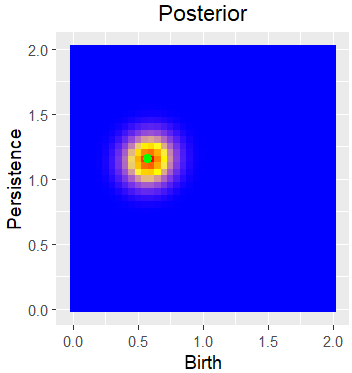}}
     \subfloat[]{\includegraphics[width=1.5in,height=1.4in]{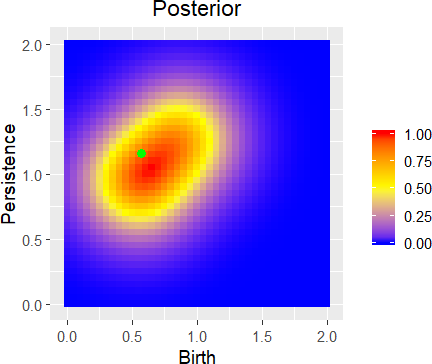}}
      
     \caption{\textit{Case-I: Posterior intensities obtained by using Proposition \ref{prop:post}. We consider informative (a), weakly informative (d), unimodal uninformative (g), and  bimodal uninformative (j) prior intensities . The color maps represent scaled intensities. The list of associated parameters of observed PD used for this case are in Table \ref{tabel:posterior parameters_1}. Posteriors computed from all of these priors estimate the $1-$dimensional hole accurately for a choice of variances in the observed persistence diagram as $\sigma^{\D_{Y_O}}=0.01\,\,\, \text{and} \,\,\,\sigma^{\D_{Y_S}}=0.1$ which are presented in (b), (e), (h) and (k). After increasing the variance to  $\sigma^{\D_{Y_O}}=0.1$, we observe the posteriors can still estimate hole with higher variance as presented in (c), (f), (i) and (k). 
     } }
    \label{fig:prior_like_noise_comp_1}
 \end{figure}

\begin{figure} [h!]
  \centering
  \hspace{0.05in}\subfloat[]{\includegraphics[width=1.3in,height=1.4in]{informative_prior.png}}
    \subfloat[]{\includegraphics[width=1.3in,height=1.4in]{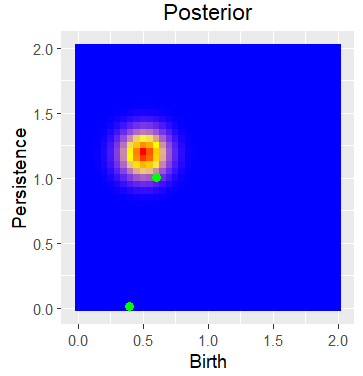}}
     \subfloat[]{\includegraphics[width=1.5in,height=1.4in]{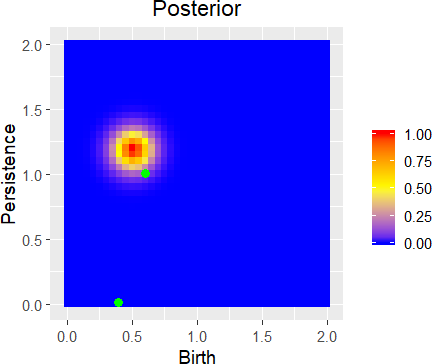}}\\ \subfloat[]{\includegraphics[width=1.3in,height=1.4in]{weakly_informative_prior.png}}
    \subfloat[]{\includegraphics[width=1.3in,height=1.4in]{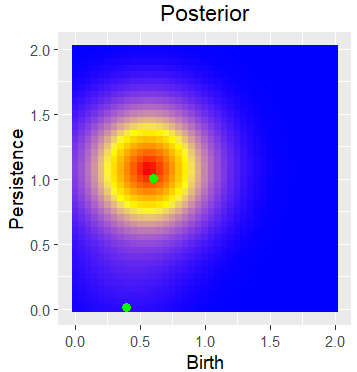}}
     \subfloat[]{\includegraphics[width=1.5in,height=1.4in]{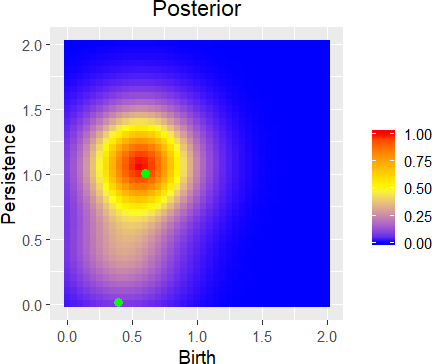}}\\
     \subfloat[]{\includegraphics[width=1.3in,height=1.4in]{uninf_prior.png}} 
     \subfloat[]{\includegraphics[width=1.3in,height=1.4in]{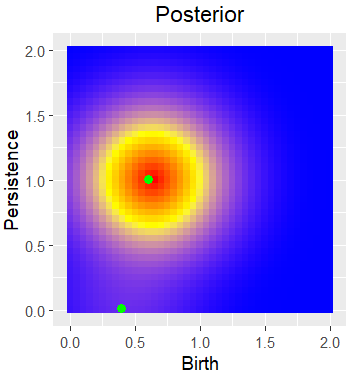}}
     \subfloat[]{\includegraphics[width=1.5in,height=1.4in]{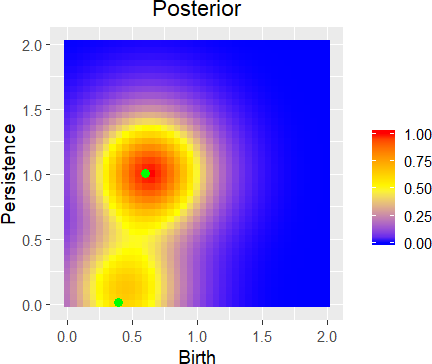}}\\
      \subfloat[]{\includegraphics[width=1.3in,height=1.4in]{bimodal_uninformative_prior.png}}
     \subfloat[]{\includegraphics[width=1.3in,height=1.4in]{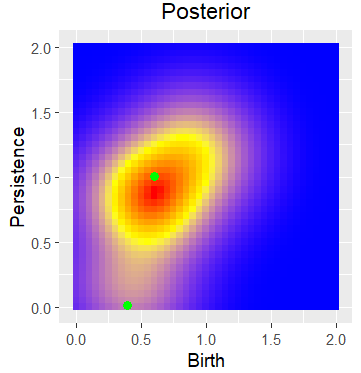}}
     \subfloat[]{\includegraphics[width=1.5in,height=1.4in]{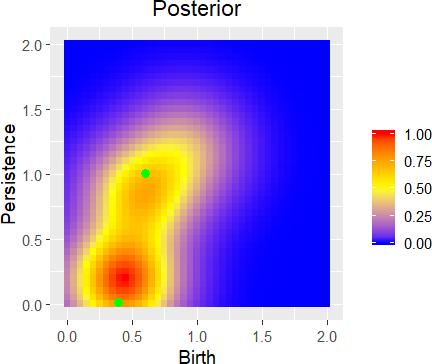}}
     \caption{\textit{Case-II:  We consider informative (a), weakly informative (d), unimodal uninformative (g), and bimodal uninformative (j) prior intensities  to estimate posterior intensities using Proposition \ref{prop:post}. The color maps represent scaled intensities. The parameters we use for estimating the posterior intensity are listed in Table \ref{tabel:posterior parameters_1}. The posterior intensities estimated from the informative prior in (b) and (c) estimate the $1-$dimensional hole with high certainty. Also, the posterior intensities estimated from the weakly informative and uninformative priors in (e), (h), and (k) imply existence of a hole with lower certainty. 
     (c), (f), (i) and (l) represent the posterior with higher variance in observed PD component  $\D_{Y_S}$. 
     As this makes the assumption that every observed point is associated to $\D_X$, we observe increased intensity skewed towards the spurious point in (f). Furthermore in (i) and (l), we observe bimodality in the posterior intensity.} }
    \label{fig:prior_like_noise_comp_2}
    \vspace{-0.2in}
 \end{figure}

\begin{figure} [h!]
  \centering
      \subfloat[]{\includegraphics[width=1.3in,height=1.4in]{informative_prior.png}}
    \subfloat[]{\includegraphics[width=1.3in,height=1.4in]{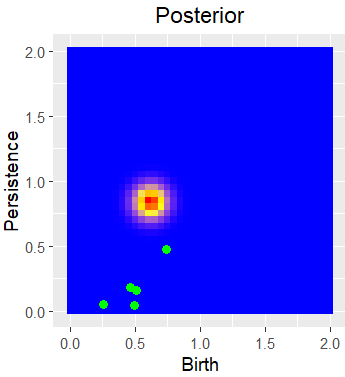}}
     \subfloat[]{\includegraphics[width=1.5in,height=1.4in]{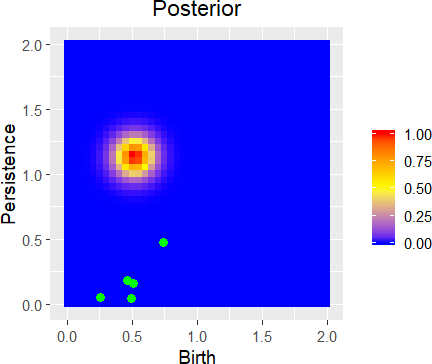}}\\
      \subfloat[]{\includegraphics[width=1.3in,height=1.4in]{weakly_informative_prior.png}}
    \subfloat[]{\includegraphics[width=1.3in,height=1.4in]{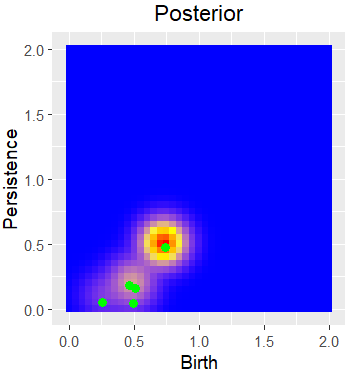}}
     \subfloat[]{\includegraphics[width=1.5in,height=1.4in]{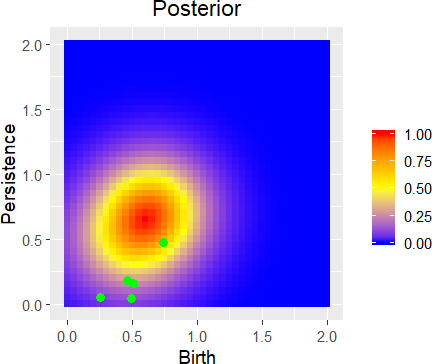}}\\
     \subfloat[]{\includegraphics[width=1.3in,height=1.4in]{uninf_prior.png}}
     \subfloat[]{\includegraphics[width=1.3in,height=1.4in]{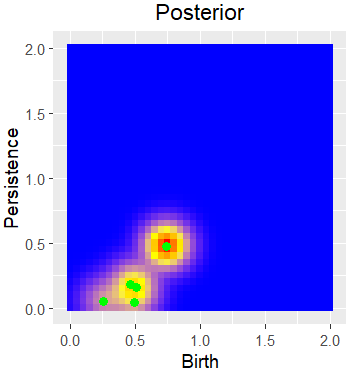}}
     \subfloat[]{\includegraphics[width=1.5in,height=1.4in]{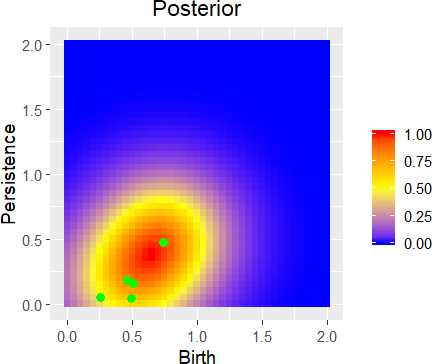}}\\
     \subfloat[]{\includegraphics[width=1.3in,height=1.4in]{bimodal_uninformative_prior.png}}
     \subfloat[]{\includegraphics[width=1.3in,height=1.4in]{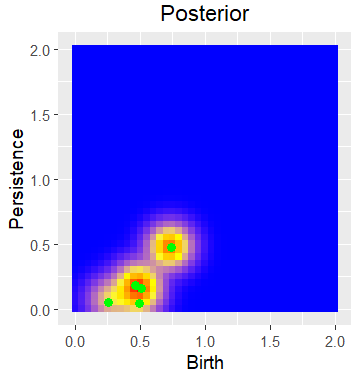}}
     \subfloat[]{\includegraphics[width=1.5in,height=1.4in]{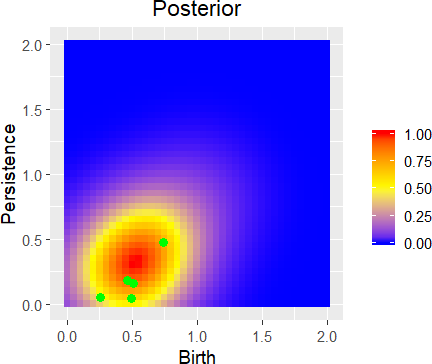}}
     \caption{\textit{Case-III: Posterior intensities obtained by using Proposition \ref{prop:post}. We consider informative (a), weakly informative (d), unimodal uninformative (g), and bimodal uninformative (j) prior intensities. The color maps represent scaled intensities. Parameters of the observed PD used to estimate the posterior intensity are listed in Table \ref{tabel:posterior parameters_1}. With a choice of $\sigma^{\D_{Y_O}} = 0.01$ and $\sigma^{\D_{Y_s}} = 0.1$, we observe the posteriors can deduce existence of the prominent feature as presented in (b), (e), (h), and (k) as we have more confidence on the component of observed data associated to prior. Otherwise, with an increased variance $\sigma^{\D_{Y_O}}=0.1$, only the posterior intensity from the informative prior is able to detect the hole with high certainty, as observed in (c). For the weakly informative and uninformative priors, the posteriors in (f), (i) and (l)  may not detect the hole directly, but the mode of (f) with higher variance and the tail towards the prominent point in (i) and (l) imply the existence of a hole in the underlying PD.}}.
    \label{fig:prior_like_noise_comp_3}
    \vspace{-0.2in}
 \end{figure}

We start by considering a Poisson PP with prior intensity $\lambda_{\mathcal{D}_X}$ that has the Gaussian mixture form given in (M2$'$). We take into account four types of prior intensities: (i) informative, (ii) weakly informative, (iii) unimodal uninformative, and (iv) bimodal uninformative; see Figures \ref{fig:prior_like_noise_comp_1}--\ref{fig:prior_like_noise_comp_3} (a), (d), (g), (j), respectively. We use one Gaussian component in each of the first three priors as the underlying shape has single $1-$dimensional feature and two for the last one to include a case where we have no information about the cardinality of the underlying true diagram. The parameters of the Gaussian mixture density in Equation \eqref{eqn:intensity of Dx} used to compute these prior intensities are listed in Table \ref{tabel:prior parameters}. To present the intensity maps uniformly throughout this example while preserving their shapes, we divide the intensities by their corresponding maxima. This ensures all intensities are on a scale from $0$ to $1$, and we call it the scaled intensity. The observed PDs are generated from point clouds sampled uniformly from the unit circle and then perturbed by varying levels of Gaussian noise; see Figure \ref{fig:circle_data} wherein we present three point clouds sampled with Gaussian noise having variances $0.001I_2$, $0.01I_2$, and $0.1I_2$, respectively. Consequently, these point clouds provide persistence diagrams $D_{Y^i}$ for $i=1,2,3$, which are considered as independent samples from Poisson point process $\D_Y$, exhibiting distinctive characteristics such as only one prominent feature with high persistence and no spurious features (\emph{Case-I}), one prominent feature with high persistence and very few spurious features (\emph{Case-II}), and one prominent feature with medium persistence and more spurious features (\emph{Case-III}). 

 \begin{table}[h!]
 \captionsetup{labelformat=empty, labelsep=newline}
 \caption{\hspace{3in} TABLE 3 \newline \footnotesize \textit{Parameters for (M3$'$) in Equation \eqref{eqn:stachastic kernel gaussian} and \eqref{eqn:intensity of Dys}. We set the weight and mean of the Gaussian component, $c^{\D_{Y_S}}=1$ and $\mu^{\D_{Y_S}}=(0.5,0)$ respectively for all of the cases. The first row corresponds to parameters in the functions characterizing $\D_Y$ that are used in computing the posterior depicted in the first column of Figure \ref{fig:prior_like_noise_comp_4}. The second row corresponds to analogous parameters that are used in computing the posterior depicted in the second columns of Figures \ref{fig:prior_like_noise_comp_1}--\ref{fig:prior_like_noise_comp_4}. Similarly, the third row corresponds to parameters in the functions characterizing $\mathcal{D}_Y$ used for computing the posterior presented in the third columns of Figures \ref{fig:prior_like_noise_comp_1}--\ref{fig:prior_like_noise_comp_4}.}\label{tabel:posterior parameters_1}}
 \begin{center}
\begin{tabular}{|c|c|c|c|}
\hline
\emph{Case-I}  & \emph{Case-II} & \emph{Case-III}  & \emph{Case-IV}\\ \hline
& & & \shortstack{$\sigma^{\D_{Y_O}}=0.1$ \\ $\sigma^{\D_{Y_S}}=0.1$} \\ \hline
\shortstack{$\sigma^{\D_{Y_O}}=0.01$ \\ $\sigma^{\D_{Y_S}}=0.1$} & \shortstack{$\sigma^{\D_{Y_O}}=0.1$ \\ $\sigma^{\D_{Y_S}}=0.1$} &\shortstack{$\sigma^{\D_{Y_O}}=0.01$ \\ $\sigma^{\D_{Y_S}}=0.1$} &\shortstack{$\sigma^{\D_{Y_O}}=0.1$ \\ $\sigma^{\D_{Y_S}}=1$} \\ \hline
\shortstack{$\sigma^{\D_{Y_O}}=0.1$ \\ $\sigma^{\D_{Y_S}}=0.1$} & \shortstack{$\sigma^{\D_{Y_O}}=0.1$ \\ $\sigma^{\D_{Y_S}}=1$} &\shortstack{$\sigma^{\D_{Y_O}}=0.1$ \\ $\sigma^{\D_{Y_S}}=0.1$} &\shortstack{$\sigma^{\D_{Y_O}}=0.01$ \\ $\sigma^{\D_{Y_S}}=0.1$}\\
\hline
\end{tabular}
\end{center}
\vspace{-0.2in}
\end{table}

For each observed PD, persistence features are presented as green dots overlaid on their corresponding posterior intensity plots. For \emph{Cases-I-III}, we set the probability $\alpha$ of the event that a feature in $\D_X$ appears in $\D_Y$ to $1$, i.e., any feature in $\D_X$ is certainly observed through a mark in $\D_Y$, and later in \emph{Case-IV}, we decrease $\alpha$ to $0.5$ while keeping all other parameters the same for the sake of comparison. The choice of $\alpha=0.5$ anticipates that any feature has equal probability to appear or disappear in the observation and in turn provides further intuition about the contribution of prior intensities to the estimated posteriors.  We observe that in all
cases, the posterior estimates the $1-$dimensional hole; however, with different uncertainty each time. For example, for the cases where the data are trustworthy, expressed by a likelihood with tight variance, or in the case of an informative prior, the posterior accurately estimates the $1-$dimensional hole. In contrast, when the data suffer from high uncertainty and the prior is uninformative, then the posterior offers a general idea that the true underlying shape is a circle, but the exact estimation of
the 1-dimensional hole is not accurate. We examine the cases below.  

\emph{Case-I:} We consider informative, weakly informative, unimodal uninformative and bimodal uninformative prior intensities as presented in Figure \ref{fig:prior_like_noise_comp_1} (a), (d), (g) and (j) respectively to compute corresponding posterior intensities. The prior intensities parameters are listed in Table \ref{tabel:prior parameters}. The observed PD is obtained from the point cloud in Figure \ref{fig:circle_data} (left). The parameters associated to the observed PD are listed in Table \ref{tabel:posterior parameters_1}.  For the observed PD arising from data with very low noise, we observe that the posterior computed from any of the priors predicts the existence of a one dimensional hole accurately.  Firstly, with a low variability in observed persistence diagram $(\sigma^{\D_{Y_O}}=0.01\,\,\, \text{and} \,\,\,\sigma^{\D_{Y_S}}=0.1)$, the posterior intensities estimate the hole with high certainty (Figure \ref{fig:prior_like_noise_comp_1} (b), (e), (h) and (k) respectively). Next, to determine the effect of observed data on the posterior, we increase the variance of the observed PD component $\D_{Y_O}$, which consists of features in observed PDs that are associated to the underlying prior. Here, we observe that the posterior intensities still estimate the hole accurately due to the trustworthy data; this is evident in Figure \ref{fig:prior_like_noise_comp_1} (c), (f), (i) and (l). In Figure \ref{fig:prior_like_noise_comp_1}, the posteriors in (b), (e), (h), and (k) have lower variance around the 1-dimensional feature in comparison to (c), (f), (i), and (l) respectively.

 \FloatBarrier
 \setcounter{table}{7}
\begin{table}[h!]
\begin{center}
     \begin{tabular}{ m{1em} m{9em} m{9em} m{12em}}

       & \hspace{0.45in}\footnotesize{\emph{Case-I}} & \hspace{0.45in} \footnotesize{\emph{Case-II}} & \hspace{0.45in}\footnotesize{\emph{Case-III}} \\ 
 \rotatebox[origin=c]{90}{\footnotesize{Informative}} & \includegraphics[width=1.25in,height=1.4in]{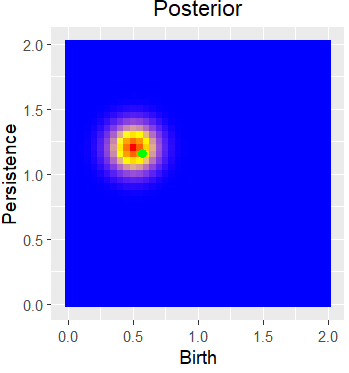}
      & 
     \includegraphics[width=1.25in,height=1.4in]{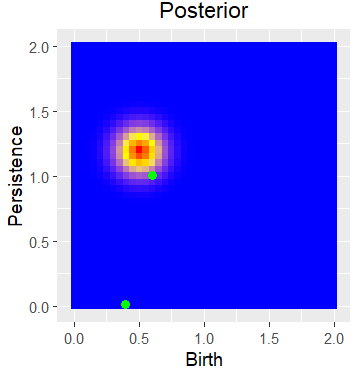}
      & 
   \includegraphics[width=1.55in,height=1.4in]{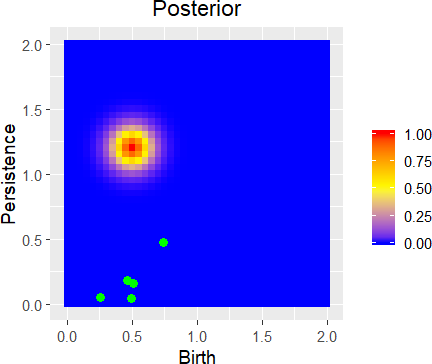}\\
     \rotatebox[origin=c]{90}{\footnotesize{Weakly Informative}} & \includegraphics[width=1.25in,height=1.4in]{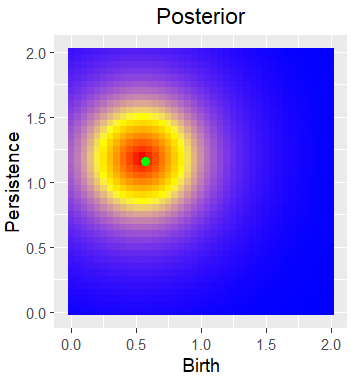}
      & 
    \includegraphics[width=1.25in,height=1.4in]{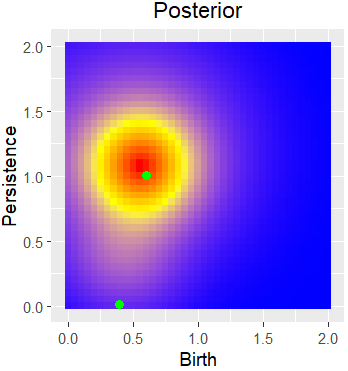}
      & 
     \includegraphics[width=1.55in,height=1.4in]{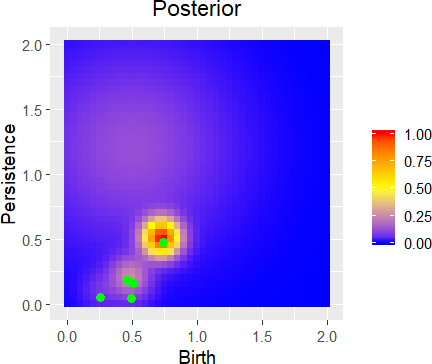}\\
     \rotatebox[origin=c]{90}{\footnotesize{Unimodal Uninformative}}& \includegraphics[width=1.25in,height=1.4in]{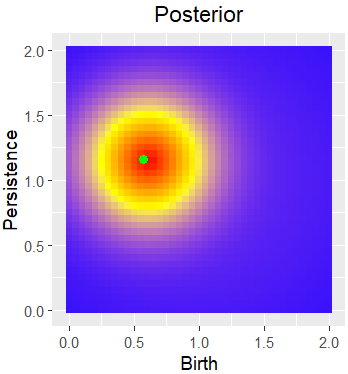}
      & 
     \includegraphics[width=1.25in,height=1.4in]{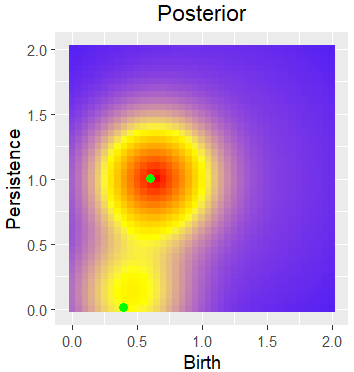}
      & 
     \includegraphics[width=1.55in,height=1.4in]{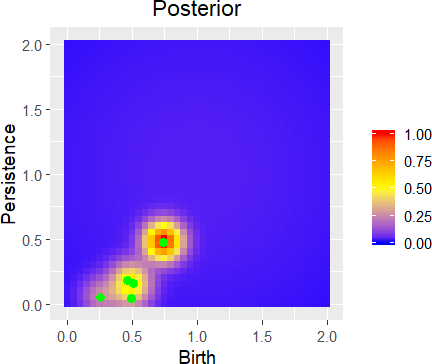}\\
     \rotatebox[origin=c]{90}{\footnotesize{Bimodal Uninformative}}& \includegraphics[width=1.25in,height=1.4in]{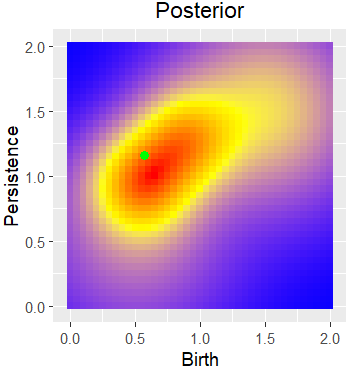}
      & 
     \includegraphics[width=1.25in,height=1.4in]{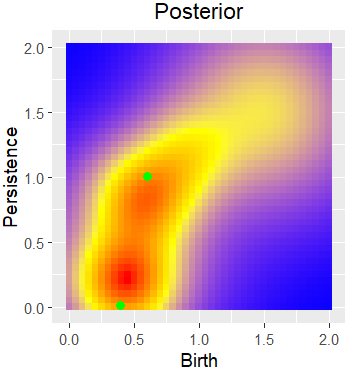}
      & 
     \includegraphics[width=1.55in,height=1.4in]{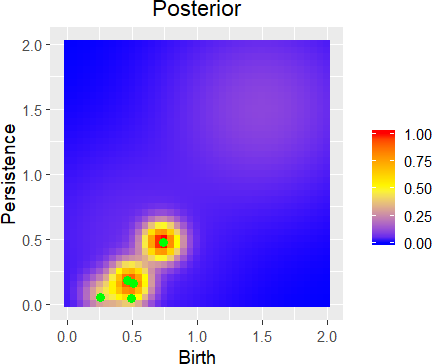}
      \end{tabular}
       \captionsetup{labelformat=empty}
      \caption{FIG. 8: \textit{Case-IV: The first, second and third columns match the parameters of observed persistence diagram $\D_Y$ used in computing the posteriors depicted in the third column of Figure \ref{fig:prior_like_noise_comp_1} and Figure \ref{fig:prior_like_noise_comp_2}, and second column of Figure \ref{fig:prior_like_noise_comp_3} respectively with $\alpha=0.5$. 
       The parameters are presented in Table \ref{tabel:posterior parameters_1}.  The color maps are representing scaled intensities. A variation in the level of intensity is observed for all of them compared to their respective cases due to the added term in the posterior intensity. The posterior intensities in first and second columns  exhibits the estimation  of hole with higher variability as compared to the respective figures in Case-I and Case-II. The posteriors in the third column demonstrate dominance of the prior relative to their corresponding figures in Case-III, especially when one examines those for informative, weakly informative and bimodal uninformative priors.}}
      \label{fig:prior_like_noise_comp_4}
      \end{center}
      \end{table}

 \emph{Case-II:}  Here, we consider all four priors as in \emph{Case-I} (see Figure \ref{fig:prior_like_noise_comp_2} (a), (d), (g) and (j)).  The point cloud in  Figure \ref{fig:circle_data} (center) is more perturbed around the unit circle than that of \emph{Case-I} (Gaussian noise with variance $0.01I_2$). Due to this, the associated PD exhibits spurious features. The parameters used for this case are listed in Table \ref{tabel:posterior parameters_1}. We compute the posterior intensities for each type of prior. First, to illustrate the posterior intensity and check the capability of detecting the $1-$dimensional feature, we use moderate noise for the observed PD  $(\sigma^{\D_{Y_O}}=0.1\,\,\, \text{and} \,\,\,\sigma^{\D_{Y_S}}=0.1)$. The results are presented in Figure \ref{fig:prior_like_noise_comp_2} (b), (e), (h), and (k); overall, the posteriors estimate the prominent feature with different variances in their respective posteriors. Next, to illustrate the effect of observed data on the posterior, we increase the variance $\sigma^{\D_{Y_S}}$ of $\D_{Y_S}$. According to our Bayesian model, the persistence diagram component $\D_{Y_S}$ contains features that are not associated with $\D_X$, so increasing $\sigma^{\D_{Y_S}}$
yields that every observed point is linked to $\mathcal{D}_X$, and therefore one may expect to observe increased intensity skewed towards the spurious points that arise from noise.
Indeed, posterior intensities with weakly informative, unimodal uninformative, and bimodal uninformative priors exhibit the skewness toward the spurious point in Figure \ref{fig:prior_like_noise_comp_2} (f), (i) and (l) respectively, but this is not the case when an informative prior is used. In (f), we observe increased intensity skewing towards the spurious points, and in (i) and (l) the intensity appears to be bimodal with two modes -- one at the prominent and other at the spurious point.  For the bimodal uninformative prior since one mode is located close to the spurious point in the observed PD, we observe higher intensity for that mode in the posterior (Figure \ref{fig:prior_like_noise_comp_2} (l)) with another mode estimating the prominent feature.   

\emph{Case-III:} We again consider the four types of priors here. The observed PD constructed from the point cloud in Figure \ref{fig:circle_data} (right). The point cloud has Gaussian noise with variance $0.1I_2$ and due to the high noise level in sampling relative to the unit circle, the associated PD exhibits one prominent feature and several spurious features. We repeat the parameter choices as in \emph{Case-I} for the variances of observed PD. For the choice of $\sigma^{\D_{Y_O}}=0.01$ and $\sigma^{\D_{Y_S}}=0.1$, the posteriors computed from all of the four priors are able to detect the difference between the one prominent and other spurious points. We increase the variance of $\D_{Y_O}$ to determine the effect of observed PD on the posterior and we observe that only the posterior intensity from informative prior has evidence of the hole (Figure \ref{fig:prior_like_noise_comp_3}(c)). For the weakly informative and uninformative priors, while the posteriors in (f), (i) and (l) may not detect the hole clearly, in (f) we observe a mode with higher variance and in (i) and (l), a tail towards the high persistence point implying presence of a hole. It should be noted that with the informative prior the posterior intensity identifies the hole closer to the mode of the prior as we increase the variance in $\sigma^{\D_{Y_O}}$.

\emph{Case-IV:} Lastly, in this case we concentrate on the effect of $\alpha$. 
The rest of the parameters used for this case remain the same and are listed in Table \ref{tabel:posterior parameters_1}.
We decrease $\alpha$ to $0.5$ to model the scenario that a feature in $\D_X$  has equal probability to appear or vanish in observed $\D_Y$.  The columns of Figure \ref{fig:prior_like_noise_comp_4} correspond to the parameters of the observed persistence diagram $\D_Y$ used in computing the posteriors depicted in the third column of Figure \ref{fig:prior_like_noise_comp_1}, third column of Figure \ref{fig:prior_like_noise_comp_2}, and second column of Figure \ref{fig:prior_like_noise_comp_3} respectively.   
 By comparing them with their respective cases, we notice a change in the intensity level in all of these due to the first term of the posterior intensity on the right hand side of Equation \eqref{eqn:mg_posterior}. Comparing with the respective figures in \emph{Case-I}, we observe that the posterior intensities are estimating the hole with higher variability for the weakly informative and unimodal uninformative priors. For bimodal prior, we observe bimodality in the posterior. Next for \emph{Case-II}, the existence of a hole is evident for informative and weakly informative priors with higher uncertainty when compared to their previous cases. The unimodal and bimodal uninformative priors lead to bimodal and trimodal posteriors, respectively.      
 We observe that the posterior resembles the prior intensity more closely when we compare them to respective figures in \emph{Case-III}. One can especially see this with the informative, weakly informative and bimodal uninformative priors, which have significantly increased intensities at the location of the modes of prior.



 \vspace{-0.1in}
 \section{Classification}\label{sec:classification}
The Bayesian framework introduced in this paper allows us to explicitly compute the posterior intensity of a PD given data and prior knowledge. This lays the foundation for supervised statistical learning methods in classification. In this section, we build a Bayes factor classification algorithm based on notions discussed in Section \ref{sec:methods} and then apply it on materials data, in particular, on measurements for spatial configurations of atoms.

We commence our classification scheme with a persistence diagram $D$ belonging to an unknown class. We assume that $D$ is sampled from a Poisson point process $\D$ in $\W$ with the prior intensity $\lambda_{\D}$ having the form in (M2$'$). Consequently, its probability density has the form
\begin{equation} \label{eqn:poisson_density}
    p_{\mathcal{D}}(D)=\frac{e^{-\lambda}}{|D|!}\prod_{d \in D}\lambda_{\D}(d)=\frac{e^{-\lambda}}{|D|!}\prod_{d \in D}\sum_{i = 1}^{N}c^{\mathcal{D}}_{i}\mathcal{N}^{*}(d;\mu^{\mathcal{D}}_{i},\sigma^{\mathcal{D}}_{i}I),
\end{equation}
where $\lambda= \int_{\W} \lambda_{\mathcal{D}}=\E(|\mathcal{D}|)$, with probability $\alpha$ as in (M2$'$).   
Next suppose we have two training sets $T_{Y} := D_{Y_{1:n}}$ and $T_{Y'} := D_{Y'_{1:m}}$ from two classes of random diagrams $\D_{Y}$ and $\D_{Y'}$, respectively. The  likelihood densities of respective classes take the form of Equation \eqref{eqn:stachastic kernel gaussian}. 
We then follow Equation \eqref{eqn:mg_posterior} to obtain the posterior intensities of $\D$ given the training sets $T_{Y}$ and $T_{Y'}$ from the prior intensities and likelihood densities.  
In particular, the corresponding posterior probability density of $\mathcal{D}$ given the training set $T_Y$ is 

\vspace{-0.3in}
\begin{equation}
 \label{eqn:poisson_posterior_density}
    p_{\D|\D_Y} (D|T_Y) = \frac{e^{-\lambda}}{|D|!}\prod_{d \in D}\lambda_{D|T_{Y}}(d)
    =\frac{e^{-\lambda}}{|D|!}\prod_{d \in D}\,\Big[(1-\alpha)\lambda_{\mathcal{D}}(d)+\frac{\alpha}{n} \sum_{y_j \in T_{Y}}\sum_{i=1}^{N} C_{i}^{d|y_j}\mathcal{N}(d;\mu_{i}^{d|y_j},\sigma_{i}^{d|y_j}I)\Big],
\end{equation}

\noindent and the posterior probability density given $T_Y'$ is given by an analogous expression.
The Bayes factor defined by 

\vspace{-0.3in}
\begin{equation} \label{eqn:bayes factor}
    BF(D)=\frac{p_{D|\mathcal{D}_Y}(D|T_{Y})}{p_{D|\mathcal{D}_{Y'}}(D|T_{Y'})}
\end{equation} 
provides the decision criterion for assigning $D$ to either $\D_{Y}$ or $\D_{Y'}$. More specifically, for a threshold $c$, $BF(D)>c$ implies that $D$ belongs to $\mathcal{D}_{Y}$ and $BF(D)<c$ implies otherwise. We summarize this scheme in Algorithm \ref{alg} .

\begin{algorithm}
\caption{Bayes Factor Classification of Persistence Diagrams}
\label{alg}
\begin{algorithmic}[1]
\State \textbf{Input 1}: Prior intensities $\lambda_{\D_{Y}}$, and $\lambda_{\D_{Y'}}$ for two classes of diagrams $\D_{Y}$ and $\D_{Y'}$ respectively; a threshold $c>0$.
\State\textbf{Input 2}: Two training sets $T_{Y}$ and $T_{Y'}$ sampled from $\D_{Y}$ and $\D_{Y'}$, respectively.
\For{$D_{Y}\,\,\, \text{and},\,\,D_{Y'}$}
	\State Compute $p_{\D|\mathcal{D}_{Y}}(D|T_{Y})$ and $p_{\D|\mathcal{D}_{Y'}}(D|T_{Y'})$.
\EndFor
\State Compute $BF(D)$ as in Equation \eqref{eqn:bayes factor}
\If{$BF(D)>c$}
	\State assign $D$ to $\mathcal{D}_{Y}$.
\Else
	\State assign $D$ to $\mathcal{D}_{Y'}$.
\EndIf
\end{algorithmic}
\end{algorithm}
\vspace{-0.2in}


\subsection{Atom Probe Tomography Data}\label{subsec:real data}

Our goal in this section is to use Algorithm \ref{alg} to classify the crystal lattice of a noisy and sparse materials dataset, where the unit cells are either Body centered cubic (BCC) or Face centered cubic (FCC); recall Figure \ref{fig:cells}. The BCC structure has a single atom in the center of the cube, while the FCC has a void in its center but has atoms on the centers of the cubes' faces (Figure \ref{fig:cells} (b-c)). However, sparsity and noise do not allow the crystal structure to be revealed. 
For high-entropy alloys, our object of interest, APT, provides the best atomic level characterization possible. Due to the sparsity and noise in the resulting data, there are only a few algorithms for successfully determining the crystal structure; see \cite{gault2012atom,moody2011lattice}. These algorithms, designed for APT data,
rely on knowing the global structure \emph{a priori} (which is not the case for High Entropy Alloys (HEAs)) and 
seek to discover
small-scale structure within a sample.

Bypassing this restriction, the neural network architecture of \cite{ziletti2018insightful} provides a way to classify the crystal structure of a noisy or sparse dataset by looking at a diffraction image. However, the authors use a test data case with very low noise or sparsity, but not both cases together, which is the representative case of the APT data.  The algorithm is also not publicly available, so a side by side comparison of our method with theirs using HEAs is not feasible.
 
It is natural to consider persistence diagrams in this setting because they distill salient information about the materials patterns with respect to connectedness and empty space (holes) within cubic unit cells, i.e we can differentiate between atomic unit cells by examining their homological features.
In particular, after storing both types of spatial configurations as point clouds, we compute their Rips filtrations (see Section \ref{sec:background}), collecting resultant 1-dimensional homological features into PDs; see Figure \ref{fig:fcc_bcc_diag}. The data set had 200 diagrams from each class. To perform classification with Algorithm \ref{alg}, we started by specifying priors for each class, $\lambda_{\mathcal{D}_{BCC}}$ and $\lambda_{\mathcal{D}_{FCC}}$. Two scenarios were considered, namely using separate priors (Prior-1 in Table \ref{tbl:pri_params}) and the same prior (Prior-2 in Table \ref{tbl:pri_params}) for both BCC and FCC classes. In particular, 
for Prior-1 we superimpose 50 PDs from each class and find the highly clustered areas by using K-means clustering. The centers of the clusters from K-means are then used as the means in Gaussian mixture priors; see Eqn. \eqref{eqn:mg_posterior}. In this manner, we produce different priors for BCC and FCC classes. On the other hand for Prior-2 we choose a flat prior with higher variance level than that of Prior-1 for both of the classes. The parameters for these two prior intensities are in Table \ref{tbl:pri_params}.    
For all cases, we set $\sigma^{\mathcal{D}_{Y_O}} = 0.1$ and $\lambda_{\mathcal{D}_{Y_S}}(x) = 5\mathcal{N}^{*}\left(x;(0,0),0.2I\right)$. We chose a relatively high weight for $\lambda_{\mathcal{D}_{Y_S}}$ because the nature of the data implied that extremely low persistence holes were rare events arising from noise.   To perform 10-fold cross validation, we partitioned PDs from both classes into training and test sets. During each fold, we took the training sets from each class, $T_{BCC}$ and $T_{FCC}$, and input them into Algorithm \ref{alg} as $T_{Y}$ and $T_{Y'}$, respectively. Next, we computed the Bayes factor  $BF(D)=\frac{p_{\D_{BCC}}(D|T_{BCC})}{p_{\D_{FCC}}(D|T_{FCC})}$ for each diagram $D$ in the test sets. After this, we used the Bayes factors to construct  receiver operating characteristic (ROC) curves and computed the resulting areas under the ROC curves (AUCs) . Finally, we used the AUCs from 10-fold cross validation to build a bootstrapped distribution by resampling 2000 times. Information about these bootstrapped distributions is summarized in Table \ref{tbl:pri_params}, 
   which shows our scoring method almost perfectly distinguishes between the BCC and FCC classes using the Bayesian framework of Section \ref{sec:methods}. Also, it exemplifies the robustness of our algorithm as two different types of priors produce near perfect accuracy. 
    
                                                

\setcounter{figure}{8}
\begin{figure}[h!]
    \centering
    \subfloat[]{\includegraphics[width=2.4in,height=1.8in]{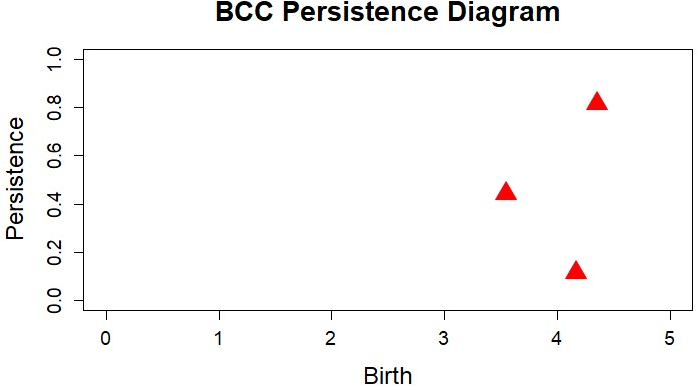}}\hspace{0.1in}
    \subfloat[]{\includegraphics[width=2.5in,height=1.9in]{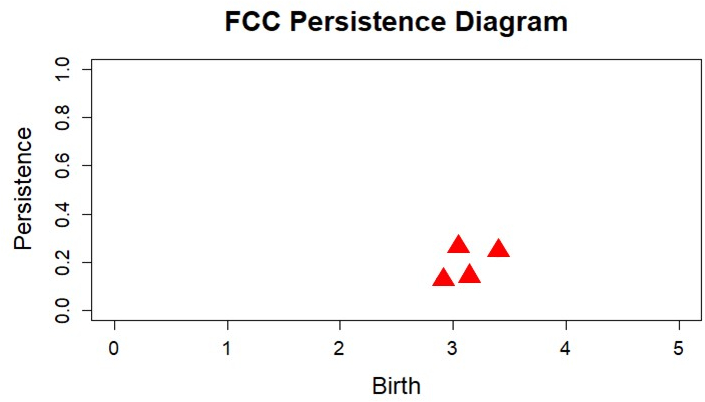}}
     \caption{ \textit{Persistence diagrams for members of the BCC and FCC classes.}}
    \label{fig:fcc_bcc_diag}
\end{figure}




    \setcounter{table}{3}

 \setcounter{table}{3}
\begin{table}[h]
	\captionsetup{labelformat=empty, labelsep=newline}
\caption{\hspace{3in} TABLE 4 \newline \footnotesize \textit{Parameters for the prior intensities used in cross-validation of materials science data. Each prior $\lambda$ is indexed by its corresponding class for Prior-1 or $U$ in the case of the Prior-2. The summary of AUCs across 10-folds for materials science data after scoring with Algorithm \ref{alg} is presented in the last three columns.}} \label{tbl:pri_params}
	\begin{center}
	{\footnotesize
	\begin{tabular}  {|m {4em}|m {3em} |m {4.5em}| m {1.5em}| m {1.5em}|m {4em}| m {3.5em}|m {4em}|} 
		
		\hline 	
		\multirow{3}{*}{\textbf{Priors}} &\multicolumn{4}{|c|}{\textbf{Parameters for Prior Intensities}} &\multicolumn{3}{|c|}{\textbf{Summary of AUC}}  \\ \cline{2-8}
		 & &\textit{$\mu_{i}^{\mathcal{D}}$} & \textit{$\sigma_{i}^{\mathcal{D}}$} & \textit{$c_{i}^{\mathcal{D}}$}&5th percentile & Mean           & 95th percentile  \\
		\hline
		\multirow{4}{*}{Prior-1} &{\textbf{$\lambda_{BCC}$}}                  & \begin{tabular}[c]{@{}l@{}}(0.5,0.24)\\ (3.6,3.6)\\(3.7,0.65)\end{tabular} & \begin{tabular}[c]{@{}l@{}}2\\ 2 \\ 2\end{tabular} & \begin{tabular}[c]{@{}l@{}}1\\ 1 \\1\end{tabular} &  \raisebox{-10.5ex}{$ 0.931$} & \raisebox{-10.5ex}{
		$0.941$} & \raisebox{-10.5ex}{$0.958$} \\ \cline{2-5}
		 & {\textbf{$\lambda_{FCC}$}}  & \begin{tabular}[c]{@{}l@{}}(0.4,0.27)\\ (2.8,1.2)\\(2.9,3)\end{tabular} & \begin{tabular}[c]{@{}l@{}}2\\ 2 \\ 2\end{tabular} & \begin{tabular}[c]{@{}l@{}}1\\ 1 \\1\end{tabular} & & &\\
		 \hline
	Prior-2             & {\textbf{$\lambda_{U}$}} & (1,1) & 20 & 1 & $0.928$ & 0.94 & $0.951$ \\
	\hline
	
	\end{tabular}
	}
	\end{center}
\end{table}	

  

\section{Discussion and Conclusions} \label{sec: conclusion}
This work is the first approach to introduce a Bayesian framework for persistent homology. This toolbox will give the opportunity to an expert to incorporate their prior belief about the data as well as analyze the data using topological data analysis methods. To that end, we introduce point processes to model random persistence diagrams. Indeed, we incorporate the prior uncertainty by modeling persistence diagrams  as Poisson point processes and noisy observations of persistence diagrams as marked Poisson PP to model the level of confidence that observations are representatives of the ground truth. 
Considering a Poisson point process, one needs to focus on the intensity of the random process. Adapting a prior intensity and a pertinent likelihood, we prove that a posterior intensity can be retrieved. It should be noted that our Bayesian model considers persistence diagrams, which are summaries of the data at hand, for defining a substitution likelihood rather than using the underlying point cloud data. 
This does not adhere to a strict Bayesian viewpoint, as we model the behavior of the persistence diagrams without considering the underlying data (materials data in our example) used to create it; however, our paradigm incorporates prior knowledge and observed data summaries to create posterior probabilities, analogous to the notion of substitution likelihood detailed in \cite{Jeffreys1961}. The general relationship between the likelihood models related to point cloud data and those of their corresponding persistence diagrams remains an important open problem. 
Furthermore we show that using Gaussian mixture conjugate family of priors. A detailed example is presented to demonstrate posterior intensities for several interesting instances resulted from varying parameters of the model. 
We establish evidence that our Bayesian framework offers update of prior uncertainty in the light of new evidence in a similar way as the standard Bayesian for random variables.
Thus, the Bayesian inference developed here can be reliably used for machine learning and data analysis techniques \emph{directly} on the space of PDs. Indeed a classification algorithm is derived and successfully applied on materials science data to assess the capability of our Bayesian framework. 
\section*{Appendix A-Proof of Theorem \ref{thm:bayes}}

\begin{proof}
 By Theorem \ref{thm:superposition}, we decompose $\lambda_{\D_X|D_{Y^{1:m}}}$ to write
\begin{equation}
   \lambda_{\D_X|D_{Y^{1:m}}} = \lambda_{\mathcal{D}_{X_V}|D_{Y^{1:m}}} + \lambda_{\mathcal{D}_{X_O}|D_{Y^{1:m}}}
                       = (1-\alpha(x))\lambda_{\mathcal{D}_X} + \lambda_{\mathcal{D}_{X_O}|D_{Y^{1:m}}}, \label{eqn1:thmpf}
\end{equation}
where the second equality follows because $\mathcal{D}_{X_V}$ is independent of $\mathcal{D}_Y$. Theorem \ref{thm:superposition} allows us to express $\lambda_{\mathcal{D}_{X_O}}$ as the average of intensity functions $\lambda_{\mathcal{D}_{X_{O}^i}}$ for $i = 1, \cdots, m$, where the $\mathcal{D}_{X_{O}^{i}}$ are independent and equal in distribution to $\mathcal{D}_{X_O}$. That is, $\lambda_{\mathcal{D}_{X_O}}=\frac{1}{m}\sum_{i=1}^{m}\lambda_{\mathcal{D}_{X_{O}^i}}$, and by conditioning we have,
\vspace{-0.2in}
 \begin{equation}\label{eqn:aux4}
  \lambda_{\DxobsDy} = \frac{1}{m}\sum_{i=1}^{m}\lambda_{\mathcal{D}_{X^i}|D_{Y^i}}.  
 \end{equation}
So to expand Equation \eqref{eqn1:thmpf} it suffices to compute $\lambda_{\mathcal{D}_{X_{O}^i}|D_{Y^i}}$ for fixed $i$. First, we express the finite PP $(\mathcal{D}_{X},\mathcal{D}_{Y})$ as a marked Poisson PP. To this end, we adopt a construction from \cite{filters_for_spp}, the augmented space $\W' := \W \cup \{\Delta\}$,
%
%
where $\Delta$ is a dummy set used for labeling points in $\mathcal{D}_{Y_S}$. 
Next, we define the random set, $\mathcal{H} = \mathcal{H}_\W \cup \mathcal{H}_{\Delta}$ such that 
\begin{equation} \label{eqn:h}
\mathcal{H} := \Big\{(x,y) \in (\mathcal{D}_{X_O},\mathcal{D}_{Y_O})\Big\} \bigcup \Big\{(\Delta,y)|y \in \mathcal{D}_{Y_S}\Big\}.
\end{equation}

One can observe that $\mathcal{H}$ is the superposition of two marked Poisson PPs $\mathcal{H}_{\W}$ and $\mathcal{H}_{\Delta}$, taking values in $\W \times \W$ and $\Delta \times \W$, respectively. Moreover, it directly follows from $(M2)$ and $(M3)(i)$ that $\mathcal{H}_{\W}$ has marginal intensity function $\alpha(x)\lambda_{\D_X}(x)$ on $\W$ and stochastic kernel density $\ell(y|x)$ while $(M3)(ii)$ shows that $\mathcal{H}_{\Delta}$ has marginal intensity function $\lambda_{D_{Y_S}}(\W)$ on $\{ \Delta \}$ with stochastic kernel density $\frac{\lambda_{D_{Y_S}}(y)}{\lambda_{D_{Y_S}}(\W)}$. By Theorem \ref{thm:marking}, the intensity functions for $\mathcal{H}_{\W}$ and $\mathcal{H}_{\Delta}$ are $\alpha(x)\lambda_{\mathcal{D}_X}(x)\ell(y|x)$ and $\lambda_{D_{Y_S}}(y)$, respectively. Hence, applying Theorem \ref{thm:superposition} to Equation \eqref{eqn:h} reveals that the intensity function for $\mathcal{H}$, $\lambda_{\mathcal{H}}$, is given by 
\begin{equation} \label{eqn:sup_intens}
    \lambda_{\mathcal{H}}(x,y) = \alpha(x)\lambda_{\mathcal{D}_X}(x)\ell(y|x)\1_{x \in \W} + \lambda_{D_{Y_S}}(y)\1_{x = \Delta}.
\end{equation}
Let $\mathcal{H}_Y := \{y : (x,y) \in \mathcal{H}\}$, $\mathcal{H}_X := \{x : (x,y) \in \mathcal{H}\}$ be the projections of $\mathcal{H}$ onto its first and second coordinates, respectively. It immediately follows from Theorem \ref{thm:mapping} that $\mathcal{H}_Y$ is a Poisson PP on $\W$ since it is the image of $\mathcal{H}$ under a projection. Therefore, by treating the first coordinates of $\mathcal{H}$ as marks, we may express $\mathcal{H}$ as a marked Poisson PP having intensity function $\lambda_{\mathcal{H}_Y}$ on $\W$ and stochastic kernel density $p(x|y)$ from $\W$ to $\W'$. Another application of Theorem \ref{thm:marking} then implies
\begin{equation} \label{eqn:sup_intens2}
    \lambda_{\mathcal{H}}(x,y) = \lambda_{\mathcal{H}_Y}(y)p(x|y).
\end{equation}
From Equations \eqref{eqn:sup_intens} and \eqref{eqn:sup_intens2}, we obtain the identity
\begin{equation} \label{eqn:p}
    p(x|y) = \frac{\alpha(x)\lambda_{\mathcal{D}_X}(x)\ell(y|x)\1_{x \in \W} + \lambda_{D_{Y_S}}(y)\1_{x=\Delta}}{\lambda_{\mathcal{H}_Y}(y)}, \hspace{2mm} \lambda_{\mathcal{H}_Y}(y) \neq 0.
\end{equation}
 Equation \eqref{eqn:p} describes the probability density of $\mathcal{H}$ at $x \in \W'$ for $y \in \W$ fixed. Substituting Equation \eqref{eqn:p} for the Janossy density in Equation \eqref{eqn:conditional_janossy} and applying Corollary \ref{cor:conditional_intensity_mpp} gives the intensity function for the point process $\mathcal{H}_X | D_{Y^i}$ whenever $\lambda_{\mathcal{H}_Y}(y) \neq  0$ for any $y \in D_{Y^i}$,
 \begin{equation} \label{eqn:intens_h}
     \lambda_{\mathcal{H}_X | D_{Y^i}}(x) = \sum_{y \in D_{Y^i}} \frac{\alpha(x)\lambda_{\mathcal{D}_X}(x)\ell(y|x)\1_{x \in \W} + \lambda_{D_{Y_S}}(y)\1_{x=\Delta}}{\lambda_{\mathcal{H}_Y}(y)}, \hspace{2mm} \lambda_{\mathcal{H}_Y}(y) \neq 0
 \end{equation}
Restricting Equations \eqref{eqn:sup_intens} and \eqref{eqn:sup_intens2} to $\W \times \W$, we obtain

$p(x|y)\lambda_{\mathcal{H}_Y}(y)=\ell(y|x)\alpha(x)\lambda_{\mathcal{D}_X}(x)$. Thus, $\ell(y|x)\alpha(x)\lambda_{\mathcal{D}_X}(x)=0$ 

whenever $\lambda_{\mathcal{H}_Y}(y)=0$, from which we conclude $\lambda_{\mathcal{H}_Y}(y)\neq 0$ a.s. . Hence, restricting Equation \eqref{eqn:intens_h} to $\W \times \W$ yields
\begin{equation} \label{eqn:aux2}
\lambda_{\mathcal{D}_{X_O}|D_{Y^i}}(x) = \sum_{y \in D_{Y^i}} \frac{\alpha(x)\lambda_{\mathcal{D}_X}(x)\ell(y|x)}{\lambda_{\mathcal{H}_Y}(y)}, \,\,\,\,\,\,\, a.s.
\end{equation}
Notice that $\mathcal{H}_{Y}$ is the same PP as $\mathcal{D}_{Y_O}\cup\mathcal{D}_{Y_S}$. Theorems \ref{thm:mapping} and \ref{thm:marking} imply that $\mathcal{D}_{Y_O}$ is a Poisson PP, and $\mathcal{D}_{Y_S}$ is a Poisson PP by (M3), so by Theorem \ref{thm:superposition}, $\lambda_{\mathcal{H}_Y} = \lambda_{\mathcal{D}_{Y_O}} + \lambda_{\mathcal{D}_{Y_S}},$ where  $\lambda_{\mathcal{D}_{Y_O}}(y) = \lambda_{(\mathcal{D}_{X_O},\mathcal{D}_{Y_O})}(\W \times y) = \int_{\W}\alpha(u)\lambda_{\mathcal{D}_{X_O}}(u)\ell(y|u)du$ by Theorem \ref{thm:marking}. Employing Equation \eqref{eqn:aux2} one gets that
\begin{equation} \label{eqn:aux3}
    \lambda_{\mathcal{D}_{X_O}|D_{Y^i}}(x) = \alpha(x)\sum_{y \in D_{Y^i}}\frac{\ell(y|x)\lambda_{\mathcal{D}_X}(x)}{\lambda_{\mathcal{D}_{Y_S}}(y)+\int_{\W}\ell(y|u)\alpha(u)\lambda_{\mathcal{D}_X}(u) du},
\end{equation}
which proves Theorem \ref{thm:bayes} after substituting into Equation \eqref{eqn1:thmpf}.
\end{proof}
\section*{Acknowledgments}
Research has been partially funded by the Army Research Office, W911NF-17-1-0313,  the National Science Foundation, MCB-1715794, and DMS-1821241, and Thor Industries/Army Research Lab, W911NF-17-2-0141.   

\vspace{-0.2in}


\bibliographystyle{siamplain}
\bibliography{bayes_tda_pp_filter,references}
\end{document}